# Advancing oxygen-enhanced MRI acquisition and modelling through dynamic ΔR1 and ΔR2* sensing

Jorge Campos Pazmiño[1,2], Amiel Beausoleil-Morrison[1], Nathalie Panov[1,3], Véronique Fortier[1,4,5], and Ives R. Levesque[1,2,5,6]

*[1]Medical Physics Unit, McGill University, Montréal, QC, Canada*

*[2]Department of Physics, McGill University, Montréal, QC, Canada*

*[3]Department of Physiology, McGill University, Montréal, QC, Canada*

*[4]Department of Medical Imaging, McGill University Health Centre and Department of Radiology, McGill University, Montréal, QC, Canada*

*[5]Gerald Bronfman Department of Oncology, McGill University, Montréal, QC, Canada*

*[6]Research Institute of the McGill University Health Centre, Montréal, QC, Canada*

October 22, 2026

* Correspondence to:

Ives R. Levesque, PhD

Medical Physics Unit, McGill University

Cedars Cancer Centre - Glen Site, DS1.9326

1001 boul. Décarie

Montréal, QC H4A 3J1

ives.levesque@mcgill.ca

514-934-1934 ext. 48105

**Abstract**

**Purpose:** To develop a multiparametric mapping framework for quantifying fat- and water-specific longitudinal and effective transverse relaxation rates, as well as changes in these parameters induced by a gas challenge.

**Theory and Methods:** The proposed framework, termed SR-Fat DESPOT (single-reference fat-water separated driven equilibrium single-pulse observation of $T_1$), combines a two-component signal model with a single-reference variable flip angle (SR-VFA) approach. The signal model explicitly incorporates gas challenge-induced changes in relaxation rates. The single-reference acquisition strategy reduces data acquisition requirements while preserving sensitivity to both baseline parameters and dynamic changes. Sequence parameters were optimized using Cramér-Rao lower bound theory. SR-Fat DESPOT was evaluated using a modified open-source abdominal numerical phantom and in two in vivo experiments: a single-participant test–retest experiment and a pilot study with six participants.

**Results:** In the abdominal numerical phantom, SR-Fat DESPOT was accurate for quantification of baseline relaxation rates with a normalized median absolute error (nMedAE) <3% for signal-to-noise ratio (SNR) >30 when comparing ground truth and estimates. SR-Fat DESPOT was also able to accurately recover median changes in relaxation rates, with differences within 0.001 $s^{-1}$ when compared to ground truth. In vivo, SR-Fat DESPOT measured changes in relaxation rates for spleen, kidneys, liver, and muscle under 100% $O_2$ breathing, consistent with reports in literature. Changes detected in the spleen and kidneys were statistically significant across participants.

**Conclusion:** SR-Fat DESPOT can quantify changes in relaxation rates induced by a gas challenge while reducing data requirements compared to more conventional approaches used in MR-oximetry literature.

## Introduction

Magnetic resonance imaging (MRI) provides a non-invasive approach for assessing tissue oxygenation and hypoxia[1]. In particular, oxygenation can be probed through measurements of the longitudinal ($R_1$) and effective transverse ($R_2^*$) relaxation rates and their variations during a gas challenge[1]. The change in $R_1$ ($\Delta R_1$) reflects variations in dissolved oxygen ($O_2$) concentration within tissue[1,2], whereas changes in $R_2^*$ ($\Delta R_2^*$) are sensitive to alterations in blood oxygenation levels[1–3]. An increase in $R_1$ ($\Delta R_1 > 0$) corresponds to an increase in dissolved $O_2$ in tissue, while $\Delta R_2^*$ is positively correlated with deoxyhemoglobin levels[1]. Furthermore, tissue regions exhibiting minimal $\Delta R_1$ response (refractory $\Delta R_1$) combined with large negative $\Delta R_2^*$ ($\Delta R_2^* < 0$) changes are indicative of hypoxia[4,5].

Measurements of single-component $R_1$ exhibit relatively low sensitivity to changes in dissolved $O_2$[6]. This limitation can be mitigated by separating water and fat contributions and measuring component-specific relaxation rates. In particular, fat-specific $R_1$ ($R_{1f}$) can improve sensitivity in tissues with high fat content, while water-specific $R_1$ ($R_{1w}$) is more suitable for tissues with low fat content[6,7]. The relative fat content of tissue can be quantified using the proton density fat fraction (PDFF)[8].

Fat DESPOT (fat-water separated driven equilibrium single pulse observation of $T_1$) is a multiparametric mapping technique that simultaneously estimates $R_{1w}$, $R_{1f}$, $R_2^*$, and PDFF using spoiled multi-echo gradient echo (MGRE) acquisitions at multiple flip angles (FAs)[7]. The original magnitude-only implementation required short echo spacing ($\Delta$TE), which imposed two MGRE datasets per FA and prolonged acquisition times[9]. A later complex signal formulation improved model fitting, relaxed $\Delta$TE constraints, and enabled acquisition of a single MGRE dataset per FA, reducing scan time and improving clinical feasibility[9].

Despite advancements in Fat DESPOT, several challenges remain. First, its in vivo sensitivity to oxygenation-induced relaxation changes has not been fully characterized. Second, its application in abdominal imaging is complicated by motion, which can degrade image quality and parameters estimation accuracy[10]. Finally, previous implementations separated $R_{1w}$ and $R_{1f}$, but did not consider fat- and water-specific $R_2^*$ ($R_{2w}^*$ and $R_{2f}^*$, respectively).

To address these limitations, we propose single reference (SR)-Fat DESPOT, a modified framework designed to directly quantify relaxation changes during a gas challenge within a breath-hold acquisition, making it suitable for abdominal imaging. The model incorporates baseline relaxation parameters ($R_{1w}$, $R_{1f}$, $R_{2w}^*$ and $R_{2f}^*$ measured during air breathing) together with differential terms describing gas challenge-induced changes ($\Delta R_{1w}$, $\Delta R_{1f}$, $\Delta R_{2w}^*$ and $\Delta R_{2f}^*$). Inspired by the single-reference variable flip angle (SR-VFA) approach for dynamic contrast-enhanced (DCE) MRI[11–13], SR-Fat DESPOT enables dynamic quantitative mapping using baseline measurements and a single additional MGRE acquisition during the stimulus, rather than repeating the full mapping protocol at each time point as in conventional MR oximetry[14–17]. We hypothesized that a single MGRE acquisition is sufficient to detect oxygenation-induced changes in tissue and blood. In addition, bipolar readout gradients were tested to reduce $\Delta$TE, increase the number of echoes within a repetition time (TR), and improve the precision of $\Delta R_{1w}$, $\Delta R_{1f}$, $\Delta R_{2w}^*$ and $\Delta R_{2f}^*$.

SR-Fat DESPOT was evaluated in a numerical abdominal phantom simulating physiologically plausible responses to 100% $O_2$ inhalation (hyperoxic challenge). In vivo, we studied the feasibility of using SR-Fat DESPOT to detect changes induced by a hyperoxic challenge in a small cohort of healthy participants. Sequence parameters were optimized using Cramér-Rao lower bound (CRB) theory.

## Theory

### Multiparametric mapping with SR-Fat DESPOT

During a gas challenge, the steady-state MGRE signal ($S_{n,p}$) in a voxel for a two-component fat-water tissue model with bipolar readout gradients can be expressed as:

$$S_{n,p} = M_0\left((1-\mathrm{PDFF})W_p e^{-(R_{2w}^*+\Delta R_{2w}^*)TE_n} + (\mathrm{PDFF})F_p e^{-(R_{2f}^*+\Delta R_{2f}^*)TE_n}\sum_{q=1}^{Q}\alpha_q e^{i2\pi\Delta f_q TE_n}\right)e^{i2\pi\psi_p TE_n}e^{(-1)^n(i\phi_p+\epsilon_p)} \quad (1)$$

$$W_p = \frac{\left(1-e^{-(R_{1w}+\Delta R_{1w})TR}\right)\sin\left(B_1^+\theta_p\right)}{1-e^{-(R_{1w}+\Delta R_{1w})TR}\cos\left(B_1^+\theta_p\right)}e^{i\phi_{w,p}} \quad (2)$$

$$F_p = \frac{\left(1 - e^{-(R_{1f}+\Delta R_{1f})TR}\right)\sin\left(B_1^+\theta_p\right)}{1 - e^{-(R_{1f}+\Delta R_{1f})TR}\cos\left(B_1^+\theta_p\right)} e^{i\phi_{f,p}} \quad (3)$$

$M_0$ is the total magnetization, PDFF the proton density fat fraction, TR the repetition time, $B_1^+$ the effect of transmit field heterogeneities, $\theta_p$ the excitation FA for the $p^{th}$ MGRE acquisitions, $TE_n$ the echo time of the $n^{th}$ echo. $W_p$ and $F_p$ are the complex water and fat signal components presented in Equations 2 and 3. $R_{1w}$, $R_{1f}$, $R_{2w}^*$, and $R_{2f}^*$ are the baseline fat- and water-specific longitudinal and effective transverse relaxation rates. $\Delta R_{1w}$, $\Delta R_{1f}$, $\Delta R_{2w}^*$, and $\Delta R_{2f}^*$ are the changes in relaxation rates induced by the gas challenge. $\phi_{w,p}$ and $\phi_{f,p}$ are the time-independent phase components of the water and fat signals that arise during RF excitation and spoiling[18,19]. These terms are allowed to vary across signal components and FAs, as the assumption of their equivalence is only valid at low FAs[19]. The sum $\sum_{q=1}^{Q} \alpha_q e^{i2\pi\Delta f_q TE_n}$ represents a fat spectrum with Q distinct resonances with relative amplitude $\alpha_q$ and chemical shift frequency $\Delta f_q$. $\psi_p$ represents the main magnetic field inhomogeneities. Finally, $\phi_p$ and $\epsilon_p$ represent phase and amplitude errors introduced by bipolar readout gradients[20–23]. $\psi_p$, $\phi_p$, and $\epsilon_p$ are allowed to vary between acquisitions with different FAs.

Baseline parameters and gas challenge-induced relaxation changes can be estimated by combining the signal model in Equation 1 with a SR-VFA scheme. Two MGRE acquisitions at FAs $\theta_{b1}$ and $\theta_{b2}$ serve as baseline measurements, for which $\Delta R_{1w}$, $\Delta R_{1f}$, $\Delta R_{2w}^*$, and $\Delta R_{2f}^*$ are set to zero. During the gas challenge, a single MGRE measurement is collected at each time frame $t$, using FA $\theta_t$, to capture oxygenation-induced changes and estimate $\Delta R_{1w}$, $\Delta R_{1f}$, $\Delta R_{2w}^*$, $\Delta R_{2f}^*$ as free parameters. Joint mapping of baseline parameters $M_0$, PDFF, $R_{1w}$, $R_{1f}$, $R_{2w}^*$, and $R_{2f}^*$; gas challenge-induced changes $\Delta R_{1w}$, $\Delta R_{1f}$, $\Delta R_{2w}^*$, $\Delta R_{2f}^*$; and $\phi_{w,p}$, $\phi_{f,p}$, $\psi_p$, $\phi_p$, $\epsilon_p$ terms can be performed with a single fit using the data from all FAs $\theta_{b1}$, $\theta_{b2}$, and $\theta_t$ (one FA for each time frame). Details of the fitting procedure are provided in the “Data processing” subsection in the Methods section.

## Methods

### Sequence parameter optimization

CRB theory was used to optimize FAs to estimate the unknown parameters in Equation 1-3. For each parameter, the noise performance of each estimate was quantified with the equivalent number of signal averages (NSA). NSA is defined as the ratio of a parameter's variance when it is the only free parameter in the model to its variance when all estimated parameters in the model are left free[24]. NSAs were calculated over discrete combinations of tissue parameters PDFF, $R_{1w}$, $R_{1f}$, $R^*_{2w}$, $R^*_{2f}$, $\Delta R_{1w}$, $\Delta R_{1f}$, $\Delta R^*_{2w}$, and $\Delta R^*_{2f}$ for a range of sequence parameters. For each set of sequence parameters, the maximum NSA ($NSA_{max}$) across all tissue combinations was determined, and optimal sequence parameters were identified by maximizing $NSA_{max}$. To assess overall performance across estimates, the aggregate metric $\sum NSA_{max}$ was used.

The proposed CRB formalism was applied to optimize SR-Fat DESPOT considering three FAs: two baseline FAs $\theta_{b1}$ and $\theta_{b2}$, and a single FA $\theta_t = \theta_{b2}$ applied to all time frames during the gas challenge. FAs were bounded to 30° to minimize deviations from the perfectly spoiled MGRE signal model due to imperfect spoiling[25,26] and slab profile effects (although these are much less prominent in 3D acquisitions compared to 2D)[27]. The constraint $\theta_t = \theta_{b2}$ was evaluated by comparing with results from an optimization allowing $\theta_t$ to deviate from $\theta_{b2}$. Details of the NSA calculations, simulation parameters (Supplementary Table S1), and Monte Carlo (MC) validation of the CRB analysis are provided in the Supplementary Information.

For numerical simulations and in vivo experiments, the first echo time ($TE_1$), echo spacing ($\Delta TE$), and TR were constrained to their minimum values to reduce acquisition time, enabling imaging within a single breath-hold.

### Evaluation of SR-Fat DESPOT with the numerical phantom

A modified open-source abdominal numerical phantom[28] was used to evaluate SR-Fat DESPOT. The phantom includes realistic anatomy with segmented organs assigned spatially varying magnetic susceptibility, $R^*_2$, PDFF, and main magnetic field inhomogeneities. The phantom was modified to incorporate baseline $R_{1w}$ and $R_{1f}$ as well as $\Delta R_{1w}$ and $\Delta R_{1f}$ induced during 100% $O_2$ inhalation[29–31]. Because literature $R_1$ values generally do not distinguish between water and fat, values used in the simulation were assumed to represent water relaxation rates ($R_1 \approx$ $R_{1w}$). We used $R_{1f} = 2.62$ $s^{-1}$ fixed across all tissues, in accordance with the common practice of treating $R_{1f}$ as a constant[32–34]. In the model, it was assumed that $R_{2w}^* = R_{2f}^* = R_2^*$ and the map provided in the numerical phantom was used for baseline values. $\Delta R_{2w}^*$ and $\Delta R_{2f}^*$ were set to 0 since no change is expected in a hyperoxic gas challenge[29,30]. Additional modifications enabled the representation of the use of bipolar or unipolar readout gradients. The bipolar scheme had shorter $\Delta TE$ and more echoes, matching the in vivo experiments (Table 1). For bipolar gradients, the phantom incorporated phase errors and amplitude modulation. The phase errors $\phi_p$ were modelled as a linear error along the read-out direction[22], while amplitude modulation was introduced by setting $\epsilon_p$ as a constant. For unipolar gradients, both $\phi_p$ and $\epsilon_p$ were set to zero. All parameters for the numerical phantom are presented in Supplementary Table S2.

SR-Fat DESPOT was evaluated in simulations at three SNR levels (low: 10, moderate: 30, and high: 50). Analyses focused on: (i) accuracy of quantitative maps relative to ground truth; (ii) accuracy of $\Delta R_{1w}$ and $\Delta R_{1f}$ compared with single-component $\Delta R_1$ to characterize changes in tissue oxygenation; and (iii) performance differences between bipolar and unipolar readouts. Synthetic data were generated using Equation 1 signal model and Table 1 sequence parameters. Complex Gaussian noise was added to achieve the target SNR, defined as the ratio between the mean magnitude signal in the spleen (evaluated in the noiseless model, first echo, and highest FA used for baseline measurements) and the noise standard deviation.

Accuracy of the estimates compared to ground truth was quantified using the normalized median absolute error (nMedAE). Water- and fat-specific parameters were evaluated only in voxels with PDFF<0.8 and PDFF>0.2, respectively, and excluding regions with $R_{1w} = 0$. Organ-level analyses were performed in the kidneys (left and right), liver, spleen, subcutaneous adipose tissue (SAT), and visceral adipose tissue (VAT) using binary masks across 10 slices covering the whole organs in the processed slices. In addition to nMedAE, organ-level results are reported using the median and interquartile range (IQR).

**In vivo experiments**

SR-Fat DESPOT was evaluated for the detection of relaxation rate changes induced by 100% $O_2$ inhalation in two experiments in human participants: "Experiment 1", a test-retest experiment, and "Experiment 2", a pilot study with six participants. The study was approved by the Research Ethics Board of the McGill University Health Centre, and all volunteers gave informed consent. All scans were performed on a 3 T system (Ingenia, Philips Healthcare) using vendor-supplied anterior and posterior body coil arrays. During the gas challenge, 100% $O_2$ was delivered with a flow rate of 15 L/min via a non-rebreathing mask (Intersurgical EcoLite™) using the room's built-in supply. Each acquisition was performed within a single expiration breath-hold, with the field of view prescribed to maximize liver coverage in the superior-inferior direction.

Experiment 1 was performed in one healthy participant (male, 32 years) across two sessions. In each session, two time courses were obtained under room air (control) and hyperoxia (100% $O_2$), with measurements collected at 0 (immediately after baseline measurements), 10, 20, and 30 minutes (Supplementary Figure S1). Sequence parameters are provided in Table 1. This experiment assessed: (i) whether median $\Delta R_{1w}$ or $\Delta R^*_{2w}$ were different from zero, and whether measurements differed between acquisitions without and with a 100% $O_2$ inhalation; (ii) the sensitivity of water-specific ($\Delta R_{1w}$ and $\Delta R^*_{2w}$) and single-component ($\Delta R_1$ and $\Delta R^*_2$) parameters; and (iii) repeatability of these metrics. Analysis was performed independently for longitudinal and effective transverse relaxation rates. Statistical testing was performed over the medians using a fixed-effects factorial linear model including condition (control vs. oxygen), time frame (0, 10, and 20), metric ($\Delta R_1$, $\Delta R_{1w}$ or $\Delta R^*_{2w}$, $\Delta R^*_2$), organ, and their interactions, with repeat acquisition treated as a blocking factor. In the analysis, we excluded the 30-minute time frames for all acquisitions because the data were corrupted by motion during the second acquisition with 100% $O_2$, despite the breath hold. Analysis was performed on five contiguous slices, and organs were automatically segmented using MRISegmenter[35]. The organs considered were spleen, kidneys (left and right), liver, and paraspinal muscles (left and right autochthonous muscles). An example of the automatic segmentation is presented in Supplementary Figure S2. We considered a significance level of 0.05 and used a Bonferroni correction for pairwise comparisons. Repeatability was quantified using the repeatability coefficient $\mathrm{RC} = 1.96\sqrt{2}\sigma_w$ where $\sigma_w$ is the within-condition standard deviation pooled across time points. For example, for $\Delta R_1$:

$$\sigma_w = \sqrt{\frac{\sum_{j=1}^{J} \frac{\left(\Delta R_{1,(1,j)} - \Delta R_{1,(2,j)}\right)^2}{2}}{J}} \qquad (4)$$

where $\Delta R_{1,(1,j)}$ and $\Delta R_{1,(2,j)}$ are repeated measurements at time point j (and $J$ total time points) for a given organ and condition.

Experiment 2 included six healthy participants (3 male, 3 female, mean age 27±4 years) who underwent a 10-minute hyperoxic challenge (100% $O_2$). The experiment assessed the feasibility of detecting oxygen-induced changes across subjects and the use of bipolar readout gradients to maximize echo sampling in breath-hold acquisitions. This cohort included the participant who took part in Experiment 1. One participant was unable to tolerate the breath-hold acquisitions and therefore did not complete the study, leaving five datasets for analysis. Sequence parameters and a schematic representation of the experiment are presented in Table 1 and Supplementary Figure S1, respectively. For each participant and organ, the median and IQR of the relaxation rates and their changes were reported, and consistency in the direction and magnitude of changes was assessed across participants. Similar to Experiment 1, we tested whether $\Delta R_{1w}$ and $\Delta R_{2w}^{*}$ measurements were different from zero and whether measurements between 0 and 10 minutes were different from each other. Statistical testing was performed over the medians using a mixed-effects linear model including time frame (0 and 10), organ (spleen, kidneys, liver, and muscle), and their interaction as fixed effects, and participant (1–5) as a random intercept. Analysis was performed independently for $\Delta R_{1w}$ and $\Delta R_{2w}^{*}$. We considered a significance level of 0.05 and used a Bonferroni correction for pairwise comparisons. Fat-specific metrics were studied in SAT of two of the participants since the relatively low adipose tissue volume in the other participants limited the ability to assess these parameters across the entire cohort.

For the in vivo experiments, SR-Fat DESPOT incorporated two single-shot turbo spin echo (SS-TSE) acquisitions to perform $B_1^{+}$ mapping using the dual angle method (DAM)[36]. These acquisitions were performed with a larger voxel size than the MGRE datasets to maintain a reasonable total scan time for breath-hold acquisition. The sequence parameters for the SS-TSE acquisitions are provided in Table 1.

**Table 1:** Sequence parameters for experiments in vivo (Experiments 1 and 2). Numerical simulations using unipolar and bipolar readout gradient use the same $TE_1$, ΔTE, number of echoes, and voxel size (acquired) as in vivo experiments. All numerical simulations use the FAs for baseline and dynamic acquisitions as in Experiment 2. Scan time is reported per frame. (PE=phase encode, AP=anterior–posterior)

| | **Experiments 1 and 2** | **Experiment 1** | **Experiment 2** |
|---|---|---|---|
| **Pulse sequence** | **2D TSE** | **3D MGRE** | **3D MGRE** |
| **TR [ms]** | 10000 | 10.3 | 11.0 |
| **$TE_1$ [ms]** | 9 | 1.2 | 1.2 |
| **ΔTE [ms]** | - | 1.5 | 0.9 |
| **# TE** | - | 6 | 10 |
| **Readout** | Single-shot | Unipolar | Bipolar |
| **rBW [Hz/pix]** | 2784.7 | 1494.3 | 1494.3 |
| **FA [°]** | 60, 120 | 5, 30, 30, 30, 30, 30 | 5, 30, 30 |
| **Ref. pulse [°]** | 180 | - | - |
| **Slice gap [mm]** | 0.5 | - | - |
| **$\theta_{b1}$ and $\theta_{b2}$ [°]** | | 5, 30 | 5, 30 |
| **$\theta_t$ [°]** | | 30 (repeated 4 times) | 30 (repeated 2 times) |
| **PE direction** | AP | AP | AP |
| **Parallel imaging** | SENSE, R=1.5 | - | - |
| **Voxel size** | | | |
| **Acquired** | 5.0×5.0×4.5 | 2.5×2.5×5.0 | 2.5×2.5×5.0 |
| **Reconstructed** | 2.4×2.4×5.0 | 2.4×2.4×4.5 | 2.4×2.4×4.5 |
| **Scan time [s]** | 23.2 | 27.3 | 25.5 |

**Data processing**

For numerical simulations and in vivo experiments, the free parameters in Equations 1–3 were estimated via non-linear least squares fitting. The parameter estimation pipeline is summarized in Figure 1. Initial guesses for the non-linear fit for PDFF, $R^*_{2w}$, $R^*_{2f}$, $\Delta\phi_{W,p}$, $\phi_{F,p}$, $\psi_p$, $\phi_p$, $\epsilon_p$, $\Delta R_{1w}$, $\Delta R_{1f}$, $\Delta R^*_{2w}$, $\Delta R^*_{2f}$ were estimated from chemical shift-encoded fat-water separation performed at each FA. Fat-water separation for unipolar and bipolar readouts were performed with graph-cut based methods[37,38]. $\phi_p$ and $\epsilon_p$ were calculated via odd- and even-echo fat-water separation to correct phase and amplitude errors to subsequently perform fat-water separation in the corrected dataset[38]. To get the initial guess for $R^*_{2w}$ and $R^*_{2f}$, the original fat-water separation algorithms were changed to perform a 2D grid search, instead of 1D search to get the joint $R^*_2$[37]. Initial guesses for $\Delta R^*_{2w}$ and $\Delta R^*_{2f}$ were calculated from the difference between baseline and challenge acquisitions. The initial guess for PDFF was calculated from the fat-water separation results for the smallest FA, using a magnitude discrimination method[39]. Initial guesses for $\phi_{w,p}$, $\phi_{f,p}$, and $\psi_p$ for each FA were obtained from the fat-water separation results. Initial guesses for $R_{1w}$ and $R_{1f}$ were obtained using DESPOT1 theory[40] with the fat and water signals from $\theta_{b1}$ and $\theta_{b2}$. Finally, the initial guess for $\Delta R_{1w}$, $\Delta R_{1f}$ were calculated with the formulas in Equations 5 and 6 (derivations for these equations are presented in Supplementary Information "Derivation of equations for initial guess for $\Delta R_{1w}$ and $\Delta R_{1f}$").

$$\Delta R_{1w} = -\frac{1}{TR}\ln\left(\frac{W_t - M_0(1-PDFF)\sin(B_1^+\theta_t)}{W_t\cos(B_1^+\theta_t) - M_0(1-PDFF)\sin(B_1^+\theta_t)}\right) - R_{1w} \quad (5)$$

$$\Delta R_{1f} = -\frac{1}{TR}\ln\left(\frac{F_t - M_0(PDFF)\sin(B_1^+\theta_t)}{F_t\cos(B_1^+\theta_t) - M_0(PDFF)\sin(B_1^+\theta_t)}\right) - R_{1f} \quad (6)$$

The normalized root mean squared error (nRMSE) was used to assess fit quality and was normalized by the magnitude of the first-echo signal acquired at FA $\theta_{b2}$.

In numerical simulations and in vivo experiments, the sensitivity of fat- and water-specific relaxation rates to changes in oxygenation was compared with that of single-component relaxation rates. Details of the fitting procedure used to estimate $M_0$, $R_1$, $R^*_2$, $\Delta R_1$, and $\Delta R^*_2$ for the single-component model are provided in the Supplementary Information section "Fitting for a single-component signal equation".

For in vivo experiments, image registration was performed to reduce inter-frame motion using a two-step approach consisting of affine registration followed by deformable registration, implemented with Advanced Normalization Tools (ANTs)[41]. All MGRE and SS-TSE datasets were registered to the MGRE dataset collected with FA $\theta_{b2}$.

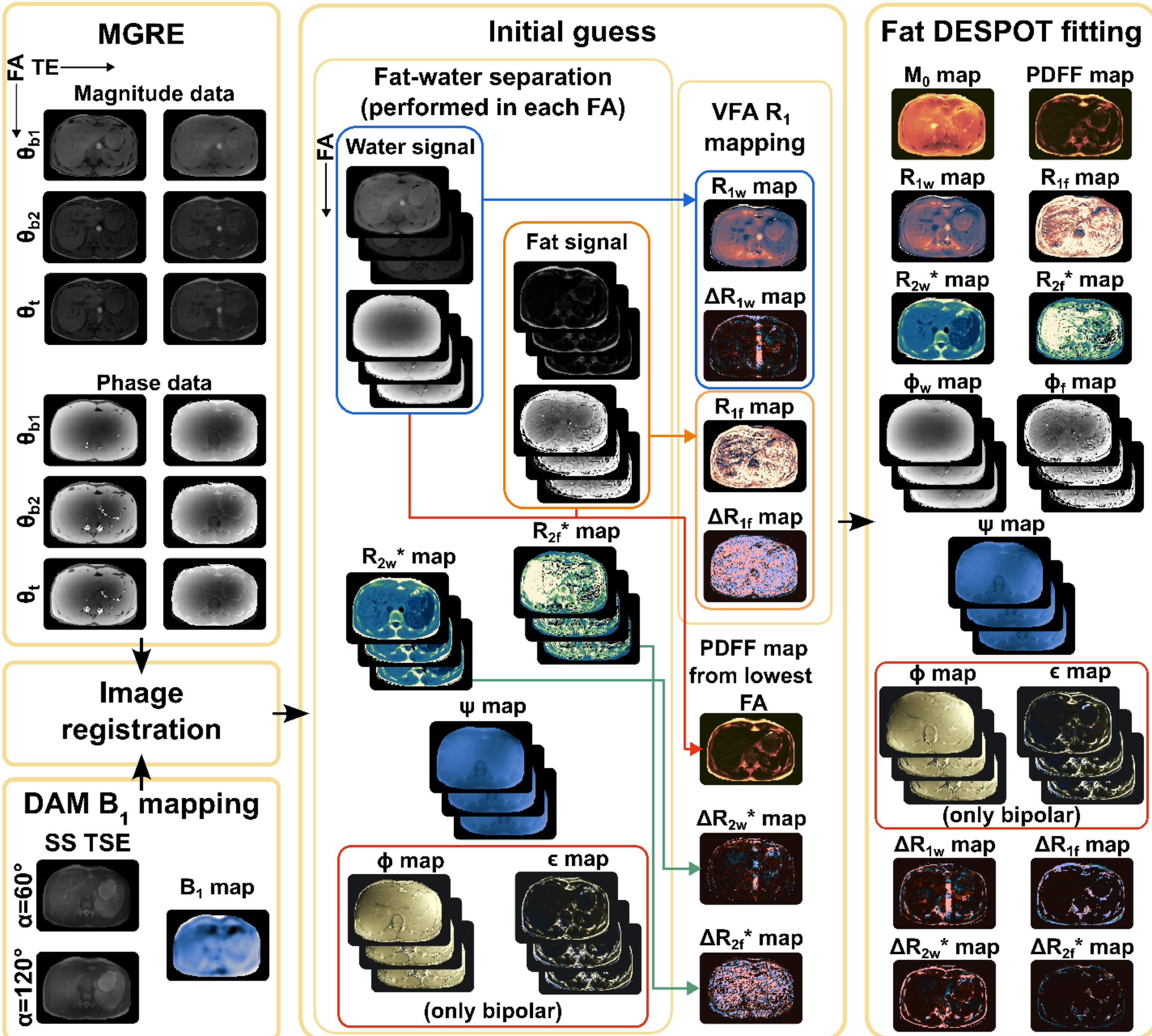

**Figure 1:** Schematic representation for the SR-Fat DESPOT framework. The measured data consists of MGRE datasets acquired at different FAs. Baseline measurements are acquired at $\theta_{b1}$ and $\theta_{b2}$, whereas the dynamic measurements at each time frame are acquired at $\theta_t$. $B_1^+$ heterogeneity is corrected via DAM $B_1^+$ mapping performed with two SS-TSE measurements. All datasets are registered using ANTs. Initial estimates for multiparametric mapping are generated by combining fat-water separation with VFA-based $R_1$ mapping. Finally, nonlinear fitting is used to estimate $M_0$, PDFF, $R_{1w}$, $R_{1f}$, $R^*_{2w}$, $R^*_{2f}$, $\Delta\phi_{w,p}$, $\phi_{f,p}$, $\psi_p$, $\phi_p$, $\epsilon_p$, $\Delta R_{1w}$, $\Delta R_{1f}$, $\Delta R^*_{2w}$, $\Delta R^*_{2f}$.

## Results

### Sequence parameter optimization

For a SR-VFA scheme with FAs $(\theta_{b1}, \theta_{b2} = \theta_t)$, we selected $(5°, 30°)$ as optimal for MR-oximetry applications. Figure 2 a) and b) show $NSA_{max}$ heat maps for multiple parameters, with the corresponding FA pair (blue marker) that maximizes performance for each. Figure 2 c) shows a heat map for $\sum NSA_{max}$. As seen in a) and b), different FA pairs optimize different estimates; thus, selection should account for which parameters are to be prioritized. The heat map in c) indicates that $(30°, 5°)$ maximizes $\sum NSA_{max}$. This pair is optimal when baseline parameter mapping is prioritized, as it provides optimal or near-optimal performance for $R_{1w}$, and $R_{1f}$. However, Figure 2 c) also shows that pairs satisfying $\theta_{b1} < \theta_{b2}$ achieve high $\sum NSA_{max}$. Under this additional constraint, the pair that maximizes $\sum NSA_{max}$ is $(5°, 30°)$. Although this choice slightly reduced noise performance for baseline parameters, it improved noise performance for $\Delta R_{1w}$, the most relevant estimate for detecting oxygenation-induced changes in water-dominant tissues.

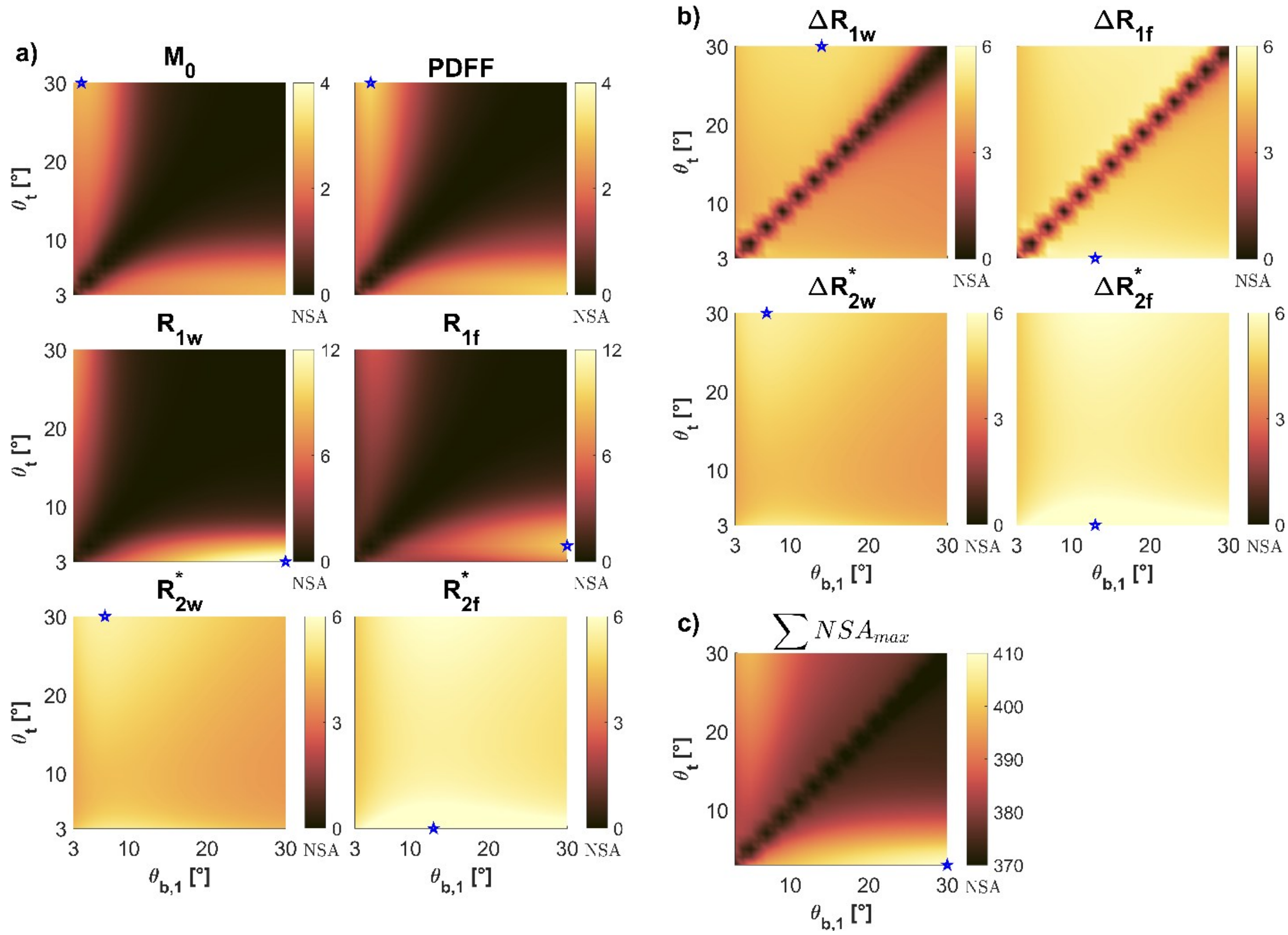

**Figure 2:** a) $NSA_{max}$ heat maps for $M_0$, PDFF, $R_{1w}$, $R_{1f}$, $R^*_{2w}$, and $R^*_{2f}$. b) $NSA_{max}$ heat maps for $\Delta R_{1w}$, $\Delta R_{1f}$, $\Delta R^*_{2w}$, and $\Delta R^*_{2f}$. $\phi_{f,p}$, $\psi_p$, $\phi_p$, $\epsilon_p$ were also included for $NSA_{max}$ calculations but are omitted from the figure. c) $\sum NSA_{max}$ heat maps. All heat maps were calculated with the constraint $\theta_{b2} = \theta_t$. The blue star marker indicates the point that maximizes $NSA_{max}$ or $\sum NSA_{max}$.

Figure S3 in Supplementary Information presents validation of the NSA calculations using MC numerical simulations. Figure S4 in Supplementary Information compares NSA across PDFF values for $(5°, 30°)$, $(30°, 5°)$ and FAs selected based on optimization that allows $\theta_t \neq \theta_{b2}$. This figure supports the selection of $(5°, 30°)$ for MR-oximetry using SR-Fat DESPOT and demonstrates that the constraint yields a better noise performance for MR-oximetry applications.

### Evaluation of SR-Fat DESPOT with the numerical phantom

In the numerical model, SR-Fat DESPOT resulted in accurate estimation of $R_{1w}$, $R_{1f}$, $R^*_{2w}$, $R^*_{2f}$, and PDFF. Figure 3 shows a comparison of the baseline maps and their ground truths for bipolar readout gradients. For SNR≥30, nMedAE<3% for all maps. Figure 3 also includes maps for $R_1$. The difference map for $R_1$ shows that even at SNR=50, $R_1$ is biased in organs like liver.

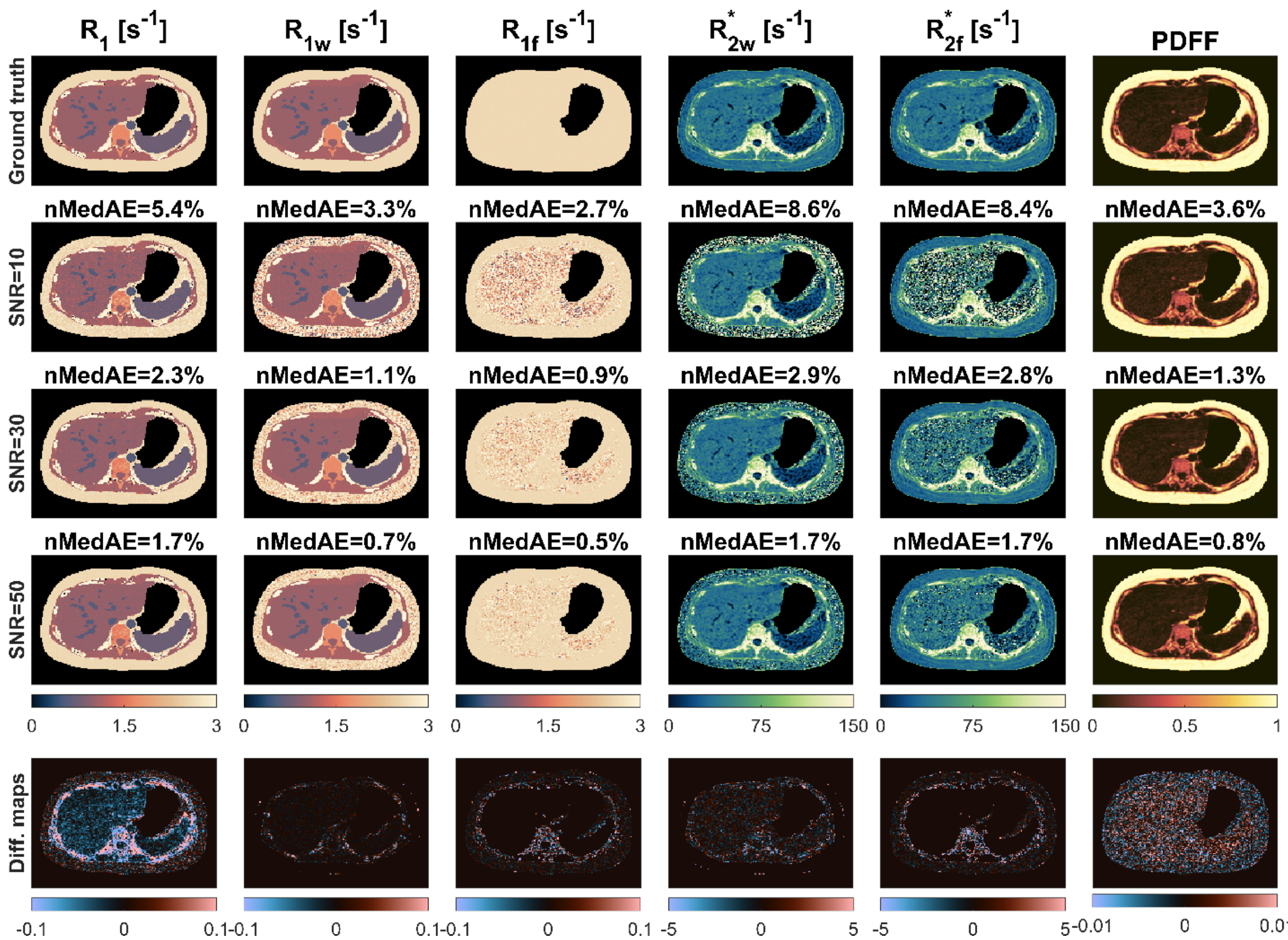

**Figure 3:** Comparison of baseline quantitative maps and ground truth for $R_1$, $R_{1w}$, $R_{1f}$, $R^*_{2w}$, $R^*_{2f}$, and PDFF using the numerical phantom including echo sampling, phase errors, and amplitude modulation effects that mimic the use of bipolar readout gradients. Rows (top to bottom) show the ground truth from the numerical simulation followed by reconstructed results at different SNR levels (10, 30, and 50). The final row presents difference maps relative to the ground truth for SNR = 50. Columns correspond to the different model parameters. The text above each map reports the nMedAE computed between the ground truth and the estimated maps.

In the simulated kidneys, liver, and spleen, SR-Fat DESPOT accurately recovered the median change in $\Delta R_{1w}$. In SAT and VAT, SR-Fat DESPOT accurately recovered the median $\Delta R_{1f}$. Table 2 reports the median and IQR values for $\Delta R_{1w}$ and $\Delta R_{1f}$. In water-dominant organs, the median $\Delta R_{1w}$ agreed with the ground truth within 0.001 $s^{-1}$, and a similar level of agreement was observed for $\Delta R_{1f}$ in SAT and VAT. Interestingly, for SNR≥30, $\Delta R_{1f}$ in kidneys, liver, and spleen converged to similar median values, albeit with larger IQRs than $\Delta R_{1w}$. However, these medians did not match the ground truth. A similar trend was observed for $\Delta R_{1w}$ in SAT and VAT. Consequently, results for $\Delta R_{1w}$ and $\Delta R_{1f}$ in fat- and water-dominant tissues are not considered for further results. Figure 4 a) shows a comparison of $\Delta R_{1w}$ and $\Delta R_{1f}$ maps and their corresponding ground truths, while Figure 4 b) presents box plots of values evaluated across organs. These maps indicate that, although SR-Fat DESPOT can recover the median changes induced by the gas challenge, voxel-wise estimates remain sensitive to noise. This sensitivity leads to higher nMedAE values than those observed for baseline parameters (Figure 3). Figure 4 further demonstrates that the noise in the maps and the spread of $\Delta R_{1w}$ and $\Delta R_{1f}$ decreases with increasing SNR, improving accuracy of voxel-wise estimates.

The average $\Delta R_1$ is an appropriate metric to describe changes in oxygenation in water- or fat-dominant tissues. Table 2 shows that median $\Delta R_1$ agreed with $\Delta R_{1w}$ and $\Delta R_{1f}$ ground truths within 0.002 $s^{-1}$. Table 2 and Figure 4 also show that IQR are smaller and $\Delta R_1$ maps are less noisy than for the equivalent $\Delta R_{1w}$ and $\Delta R_{1f}$ maps. However, in structures with intermediate fat content, $\Delta R_1$ can lead to important errors in quantification. For example, the middle of the kidney and the vertebral body were modelled with $\Delta R_{1w} > 0$ $s^{-1}$ and $\Delta R_{1w} = 0$ $s^{-1}$, respectively. However, $\Delta R_1$ depicts negative changes in these regions.

The simulated bipolar readout gradients (which had more echoes and short $\Delta$TE) resulted in more accurate and precise maps, compared to simulated unipolar readout gradients. Table 2, and Figures S5 and S6 show results for unipolar gradients. For all estimates, nMedAE and IQR were equal or lower when using simulated bipolar readout gradients.

The median $\Delta R^*_{2w}$ and $\Delta R^*_{2f}$ estimates closely matched the ground-truth values defined in the numerical phantom. Supplementary Figure S7 presents the reconstructed $\Delta R^*_{2w}$ and $\Delta R^*_{2f}$ maps for different SNR levels and readout gradient schemes, along with violin plots summarizing these parameters across multiple organs. The violin plots show that $\Delta$TE selection enabled by bipolar readout gradients produced median estimates that were closer to the ground truth and exhibited slightly improved precision compared with unipolar gradients. However, $\Delta R^*_{2w}$ and $\Delta R^*_{2f}$ estimates presented greater variability than the corresponding $\Delta R_{1w}$ and $\Delta R_{1f}$ measurements.

**Table 2:** Median and IQR values for $\Delta R_1$, $\Delta R_{1w}$, and $\Delta R_{1f}$ for calculations in the numerical phantom using unipolar and bipolar readout gradient pulses and evaluated in multiple organs.

| **Kidneys (left and right)** | | | | | |
|---|---|---|---|---|---|
| | **Readout** | **SNR=10** | **SNR=30** | **SNR=50** | **Ground truth** |
| $\Delta R_1$[s$^{-1}$] | Unipolar | 0.043 (0.014) | 0.043 (0.006) | 0.043 (0.005) | 0.048 (0.001) |
| | Bipolar | 0.043 (0.014) | 0.044 (0.006) | 0.044 (0.005) | |
| $\Delta R_{1w}$ [s$^{-1}$] | Unipolar | 0.044 (0.069) | 0.045 (0.024) | 0.046 (0.015) | 0.047 (0.001) |
| | Bipolar | 0.047 (0.058) | 0.047 (0.020) | 0.047 (0.012) | |
| $\Delta R_{1f}$ [s$^{-1}$] | Unipolar | 0.009 (0.359) | 0.035 (0.213) | 0.039 (0.141) | 0.070 (0.001) |
| | Bipolar | 0.011 (0.362) | 0.041 (0.209) | 0.042 (0.135) | |
| **Liver** | | | | | |
| | | **SNR=10** | **SNR=30** | **SNR=50** | **Ground truth** |
| $\Delta R_1$[s$^{-1}$] | Unipolar | 0.020 (0.010) | 0.020 (0.004) | 0.020 (0.003) | 0.020 (0.002) |
| | Bipolar | 0.020 (0.010) | 0.020 (0.004) | 0.020 (0.003) | |
| $\Delta R_{1w}$ [s$^{-1}$] | Unipolar | 0.023 (0.073) | 0.022 (0.024) | 0.022 (0.015) | 0.022 (0.001) |
| | Bipolar | 0.023 (0.058) | 0.022 (0.019) | 0.022 (0.012) | |
| $\Delta R_{1f}$ [s$^{-1}$] | Unipolar | 0.007 (0.359) | 0.044 (0.224) | 0.049 (0.147) | 0.070 (0.001) |
| | Bipolar | -0.003(0.367) | 0.039 (0.223) | 0.044 (0.147) | |
| **Spleen** | | | | | |
| | | **SNR=10** | **SNR=30** | **SNR=50** | **Ground truth** |
| $\Delta R_1$[s$^{-1}$] | Unipolar | 0.057 (0.014) | 0.057 (0.005) | 0.057 (0.004) | 0.060 (0.001) |
| | Bipolar | 0.057 (0.014) | 0.057 (0.005) | 0.057 (0.003) | |
| $\Delta R_{1w}$ [s$^{-1}$] | Unipolar | 0.057 (0.066) | 0.059 (0.021) | 0.059 (0.013) | 0.059 (0.001) |
| | Bipolar | 0.059 (0.055) | 0.059 (0.018) | 0.059 (0.011) | |
| $\Delta R_{1f}$ [s$^{-1}$] | Unipolar | 0.017 (0.359) | 0.019 (0.237) | 0.029 (0.171) | 0.070 (0.001) |
| | Bipolar | -0.007(0.361) | 0.020 (0.245) | 0.025 (0.174) | |
| **SAT and VAT** | | | | | |
| | | **SNR=10** | **SNR=30** | **SNR=50** | **Ground truth** |
| $\Delta R_1$[s$^{-1}$] | Unipolar | 0.070 (0.017) | 0.070 (0.006) | 0.070 (0.003) | 0.070 (0.001) |
| | Bipolar | 0.070 (0.017) | 0.070 (0.006) | 0.070 (0.003) | |
| $\Delta R_{1w}$ [s$^{-1}$] | Unipolar | 0.029 (0.339) | 0.034 (0.216) | 0.042 (0.157) | 0.070 (0.001) |
| | Bipolar | 0.018 (0.353) | 0.032 (0.220) | 0.038 (0.148) | |
| $\Delta R_{1f}$ [s$^{-1}$] | Unipolar | 0.067 (0.087) | 0.068 (0.029) | 0.069 (0.018) | 0.070 (0.001) |
| | Bipolar | 0.070 (0.068) | 0.069 (0.023) | 0.069 (0.014) | |

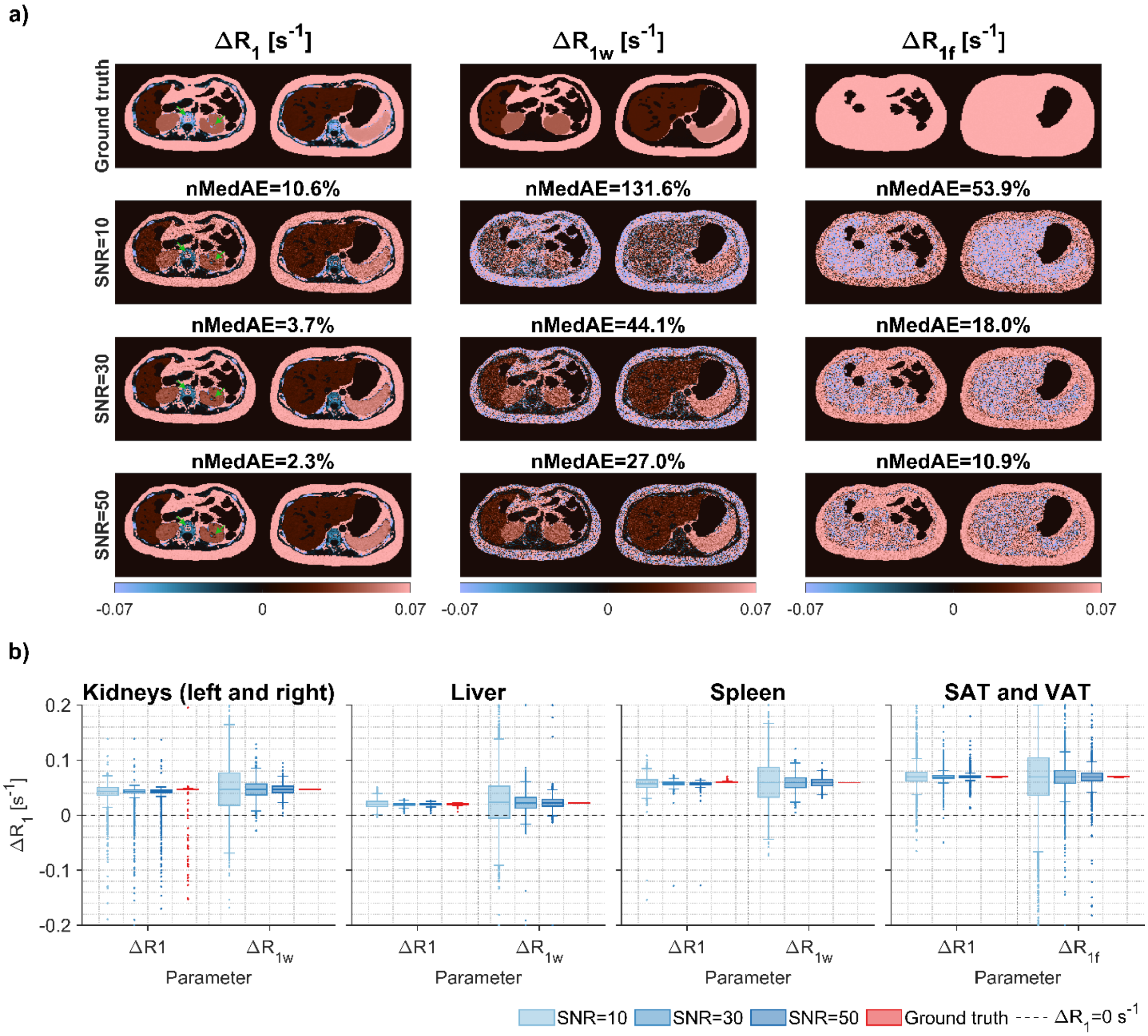

**Figure 4:** a) Comparison of quantitative maps for changes induced by a hyperoxic breathing gas challenge (100% $O_2$ inhalation) and ground truth for $\Delta R_1$, $\Delta R_{1w}$ and $\Delta R_{1f}$ using the numerical phantom including phase errors and amplitude modulation effects that mimic the use of bipolar readout gradients. The text above each map reports the nMedAE computed between the ground truth and the estimated maps. Green arrows point errors in the $\Delta R_1$ maps. b) Box plots summarizing $\Delta R_1$, $\Delta R_{1w}$ and $\Delta R_{1f}$ evaluated in masks covering the kidneys (left and right), liver, spleen, and SAT and VAT. Box plots display the median and quartiles, while the whiskers extend to 1.5 times the interquartile range. Results are presented at different SNR levels (SNR=10, 30, and 50).

**In vivo experiments**

Experiment 1 showed organ-dependent $\Delta R_{1w}$ responses to 100% $O_2$ inhalation, with the most consistent response observed in the spleen. Figure 5 a) presents quantitative maps of baseline $R_{1w}$ and $\Delta R_{1w}$, and box plots of $\Delta R_{1w}$ evaluated in multiple organs for the two time courses in one of the repetitions in Experiment 1. Supplementary Figure S8 a) and b) show results for the two repetitions of the two time courses. Table 3 reports median and IQR for $R_{1w}$ and $\Delta R_{1w}$ values measured at 10 minutes. In Experiment 1, measured $\Delta R_{1w}$ in the spleen was significantly different from zero during 100% $O_2$ inhalation at 10 and 20 minutes ($p<0.001$). In the kidneys, measured $\Delta R_{1w}$ was significantly different from zero during 100% $O_2$ inhalation at 10 minutes only ($p=0.002$). In the liver, $\Delta R_{1w}$ was significantly different from zero in experiments without and with 100% $O_2$ at 0 minutes ($p<0.001$ for both cases) and 10 minutes ($p=0.002$ and $p<0.001$, respectively). No significant differences were detected in muscle. When comparing measurements with and without the 100% $O_2$ challenge, significant differences were detected only for the spleen for the measurements at 10 and 20 minutes ($p<0.001$ in both cases). There were no statistically significant differences between $\Delta R_{1w}$ and $\Delta R_1$.

Experiment 1 showed limited $\Delta R^*_{2w}$ responses to 100% $O_2$ inhalation, with significant changes detected only in the liver. Figure 5 b) presents baseline, $\Delta R^*_{2w}$ quantitative maps, and box plots for $\Delta R^*_{2w}$ evaluated in multiple organs. Supplementary Figure S8 c) and d) show results for the two repetitions of the two time courses. Table 3 reports median and IQR for $\Delta R^*_{2w}$ values measured at 10 minutes. In Experiment 1, in the liver, measured $\Delta R^*_{2w}$ were significantly different from zero in experiments without and with 100% $O_2$ at 0, 10, and 20 minutes ($p<0.002$ for all cases). There were no statistically significant differences for any other organ, when comparing measurements without and with the 100% $O_2$ challenge, or when comparing $\Delta R^*_{2w}$ and $\Delta R^*_2$.

During air breathing, the RC measured for Experiment 1 for the spleen, kidneys, liver, and muscle were 0.058 $s^{-1}$, 0.027 $s^{-1}$, 0.047 $s^{-1}$, and 0.029 $s^{-1}$ for $\Delta R_{1w}$ and 14.4 $s^{-1}$, 3.5 $s^{-1}$, 2.0 $s^{-1}$, and 1.2 $s^{-1}$ for $\Delta R^*_{2w}$. For Experiment 1, the median of $\Delta R_{1w}$ for the spleen was consistently higher than the RC measured for this organ, which indicates that the response elicited in the spleen by a hyperoxic gas challenge is detectable beyond measurement noise.

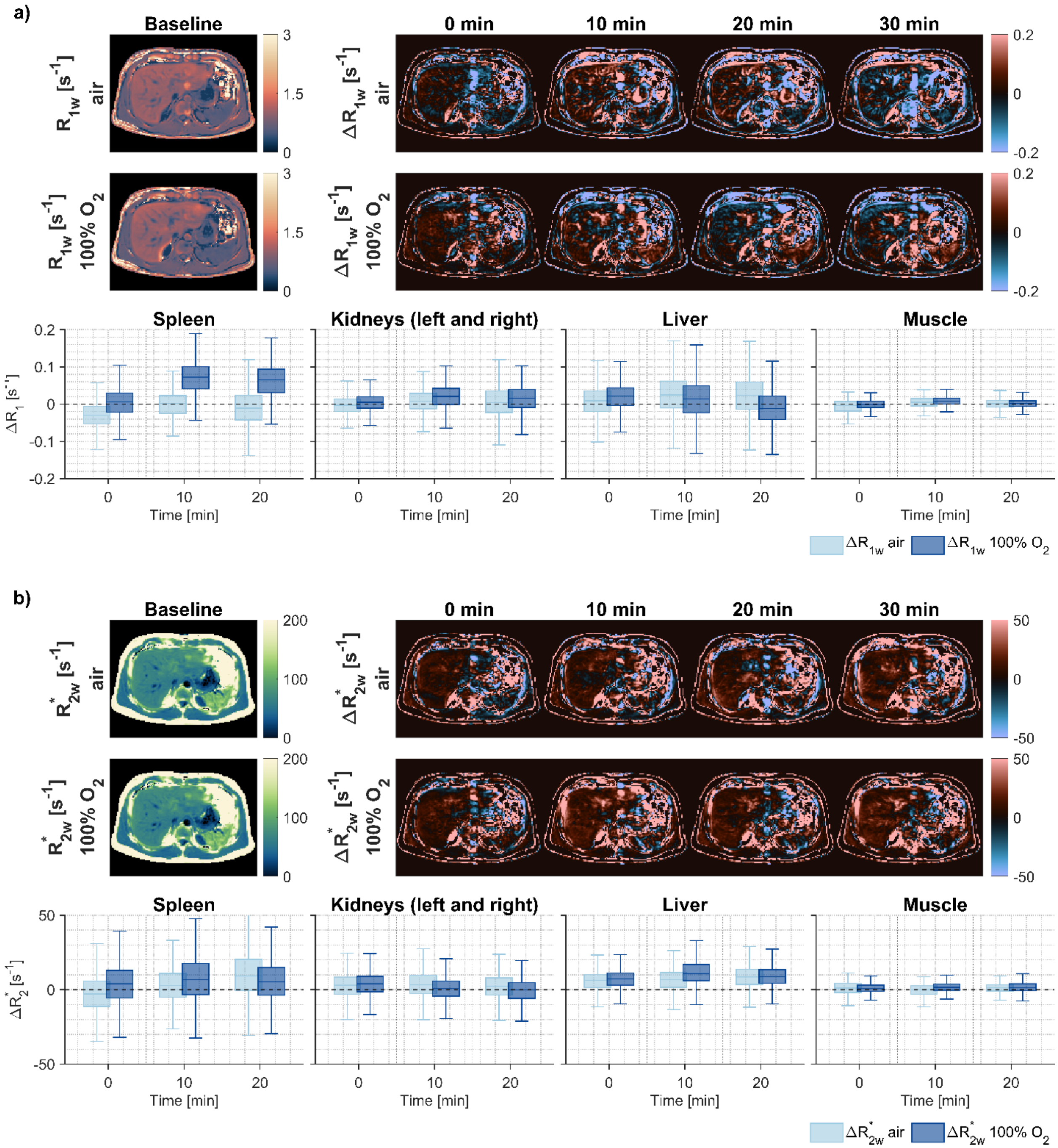

**Figure 5:** Results from one repetition of Experiment 1 (test-retest experiment). a) Quantitative maps and box plots for $R_{1w}$ and $\Delta R_{1w}$ at 0, 10, 20, and 30 minutes for two time courses acquired without and with a breathing gas challenge (100% $O_2$). b) Quantitative maps and box plots for $R^*_{2w}$ and $\Delta R^*_{2w}$. Box plots display the median and quartiles, while the whiskers extend to 1.5 times the interquartile range.

**Table 3:** Median and IQR for $R_{1w}$, $\Delta R_{1w}$, $R_{2w}^*$, and $\Delta R_{2w}^*$ measured in vivo across all participants in spleen, kidneys, liver, and muscle. $R_{1f}$, $\Delta R_{1f}$, $R_{2f}^*$, and $\Delta R_{2f}^*$ in SAT evaluated for a subset of participants (1 and 5). All values are reported for the time frame corresponding to 10 minutes during 100% $O_2$ inhalation. IQR are reported in parentheses. All values are in $s^{-1}$.

| | Spleen | | | | Kidneys (left and right) | | | |
|---|---|---|---|---|---|---|---|---|
| **Participant** | $\mathbf{R_{1w}}$ | $\mathbf{\Delta R_{1w}}$ | $\mathbf{R_{2w}^*}$ | $\mathbf{\Delta R_{2w}^*}$ | $\mathbf{R_{1w}}$ | $\mathbf{\Delta R_{1w}}$ | $\mathbf{R_{2w}^*}$ | $\mathbf{\Delta R_{2w}^*}$ |
| **#1 (Exp. 1)** | 0.581 (0.045) | 0.073 (0.060) | 108.0 (15.4) | 6.8 (20.8) | 0.476 (0.103) | 0.021 (0.043) | 17.3 (7.6) | 0.8 (10.0) |
| **#1 (Exp. 1)** | 0.553 (0.043) | 0.073 (0.054) | 102.5 (13.5) | -1.8 (16.4) | 0.438 (0.085) | 0.045 (0.035) | 17.2 (7.9) | 5.2 (11.2) |
| **#1 (Exp. 2)** | 0.571 (0.049) | 0.086 (0.035) | 116.6 (15.5) | 12.1 (13.8) | 0.466 (0.103) | 0.033 (0.036) | 18.9 (5.7) | 0.8 (9.5) |
| **#2** | 0.567 (0.075) | 0.059 (0.028) | 15.5 (3.0) | 1.6 (5.8) | 0.523 (0.109) | 0.027 (0.037) | 22.6 (14.3) | 3.0 (11.8) |
| **#3** | 0.536 (0.047) | 0.058 (0.015) | 12.7 (1.9) | -3.0 (3.2) | 0.450 (0.107) | 0.005 (0.035) | 18.1 (8.5) | -2.2 (7.1) |
| **#4** | 0.581 (0.045) | 0.060 (0.020) | 26.0 (2.9) | -2.2 (4.5) | 0.455 (0.084) | 0.016 (0.039) | 22.6 (11.3) | 3.0 (11.1) |
| **#5** | 0.524 (0.046) | 0.032 (0.017) | 18.6 (4.0) | -2.5 (4.8) | 0.377 (0.112) | 0.052 (0.051) | 14.4 (4.2) | 2.7 (7.7) |

| | Liver | | | | Muscle | | | |
|---|---|---|---|---|---|---|---|---|
| **Participant** | $\mathbf{R_{1w}}$ | $\mathbf{\Delta R_{1w}}$ | $\mathbf{R_{2w}^*}$ | $\mathbf{\Delta R_{2w}^*}$ | $\mathbf{R_{1w}}$ | $\mathbf{\Delta R_{1w}}$ | $\mathbf{R_{2w}^*}$ | $\mathbf{\Delta R_{2w}^*}$ |
| **#1 (Exp. 1)** | 0.972 (0.123) | 0.014 (0.073) | 63.5 (11.0) | 10.8 (10.8) | 0.719 (0.046) | 0.008 (0.016) | 39.8 (4.0) | 1.5 (4.1) |
| **#1 (Exp. 1)** | 0.934 (0.109) | 0.042 (0.055) | 63.2 (9.0) | 11.0 (9.4) | 0.664 (0.049) | 0.023 (0.016) | 37.6 (3.8) | 3.4 (6.1) |
| **#1 (Exp. 2)** | 0.977 (0.141) | 0.022 (0.050) | 65.5 (13.0) | 6.7 (9.0) | 0.731 (0.044) | 0.017 (0.017) | 37.6 (3.1) | 1.0 (3.7) |
| **#2** | 1.081 (0.184) | 0.020 (0.033) | 51.7 (5.2) | 3.7 (4.3) | 0.722 (0.053) | 0.007 (0.017) | 41.0 (3.1) | -2.3 (4.1) |
| **#3** | 0.832 (0.114) | -0.005 (0.035) | 36.4 (3.6) | -3.8 (4.2) | 0.694 (0.104) | -0.001 (0.012) | 35.8 (2.2) | 0.03 (2.9) |
| **#4** | 1.000 (0.159) | -0.004 (0.051) | 53.3 (5.0) | 1.5 (5.6) | 0.770 (0.075) | -0.005 (0.012) | 40.5 (3.4) | -2.5 (3.4) |
| **#5** | 0.895 (0.142) | -0.022 (0.058) | 45.1 (3.6) | -7.2 (6.0) | 0.762 (0.082) | -0.001 (0.022) | 37.5 (3.4) | 1.8 (5.5) |

| | SAT | | | |
|---|---|---|---|---|
| **Participant** | $\mathbf{R_{1f}}$ | $\mathbf{\Delta R_{1f}}$ | $\mathbf{R_{2f}^*}$ | $\mathbf{\Delta R_{2f}^*}$ |
| **#1 (Exp. 2)** | 2.061 (0.279) | -0.039 (0.072) | 38.4 (6.4) | -3.3 (2.7) |
| **#5** | 2.303 (0.260) | -0.148 (0.092) | 39.9 (10.8) | -2.3 (6.4) |

When comparing unipolar and bipolar readout gradient pulses for baseline parameter quantification, bipolar readouts yielded more robust quantification. Figure 6 presents maps acquired from the same participant on the same day using both readout strategies. Notably, PDFF maps derived from unipolar readouts exhibited a spurious left-to-right gradient, appearing as a shading artifact that artificially increased PDFF values. This artifact was not observed in the maps obtained with bipolar readouts. The corresponding reduction in nRMSE is consistent with the elimination of this shading artifact. Regarding sensitivity to oxygenation-induced changes, the use of bipolar readout gradients resulted in smaller IQR for both $\Delta R_{1w}$ and $\Delta R_{2w}^{*}$ (Table 3, Participant #1).

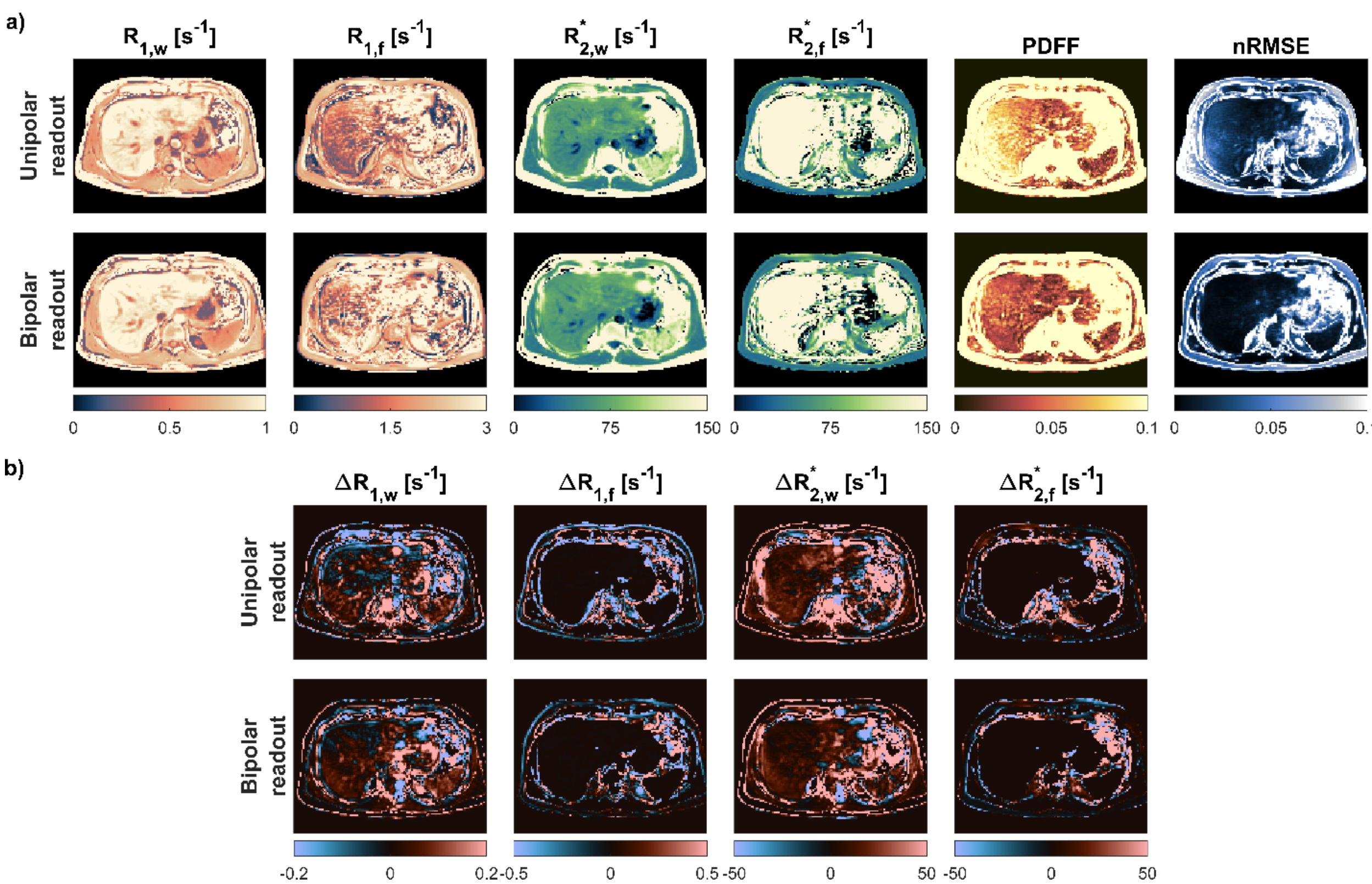

**Figure 6:** Comparison of SR-Fat DESPOT using unipolar and bipolar readout gradient pulses. Measurements correspond to the same healthy participant undergoing a hyperoxic gas challenge. All measurements were performed in a single day. a) Baseline maps and nRMSE map. b) Changes induced by the breathing gas challenge.

In Experiment 2, across participants, measured $\Delta R_{1w}$ values in the spleen and kidneys were significantly different from zero. In these organs, $\Delta R_{1w}$ also differed significantly between the 0 and 10 minute acquisitions. Model-based comparisons did not show any other statistically significant effects in the remaining organs or for the $R^*_{2w}$ analysis. Table 3 reports the median $R_{1w}$, $\Delta R_{1w}$, $R^*_{2w}$, and $\Delta R^*_{2w}$ evaluated in multiple organs for all participants.

Quantification of $\Delta R_{1f}$ and $\Delta R^*_{2f}$ was limited by variability in adipose tissue volume across participants in Experiment 2. In individuals with moderate adipose tissue, motion and ringing artifacts further precluded reliable measurement. Figure 7 illustrates three participants with differing amounts of SAT and VAT. Median and IQR values for participants #1 and #5 are summarized in Table 3. Notably, median $\Delta R_{1f}$ and $\Delta R^*_{2f}$ in SAT were negative for both participants, at time frame 0 minutes ($\Delta R_{1f}$: -0.052 $s^{-1}$ and -0.119 $s^{-1}$; $\Delta R^*_{2f}$: -4.8 $s^{-1}$ and -6.6 $s^{-1}$) and 10 minutes ($\Delta R_{1f}$: -0.039 $s^{-1}$ and -0.148 $s^{-1}$; $\Delta R^*_{2f}$: -3.3 $s^{-1}$ and -2.3 $s^{-1}$).

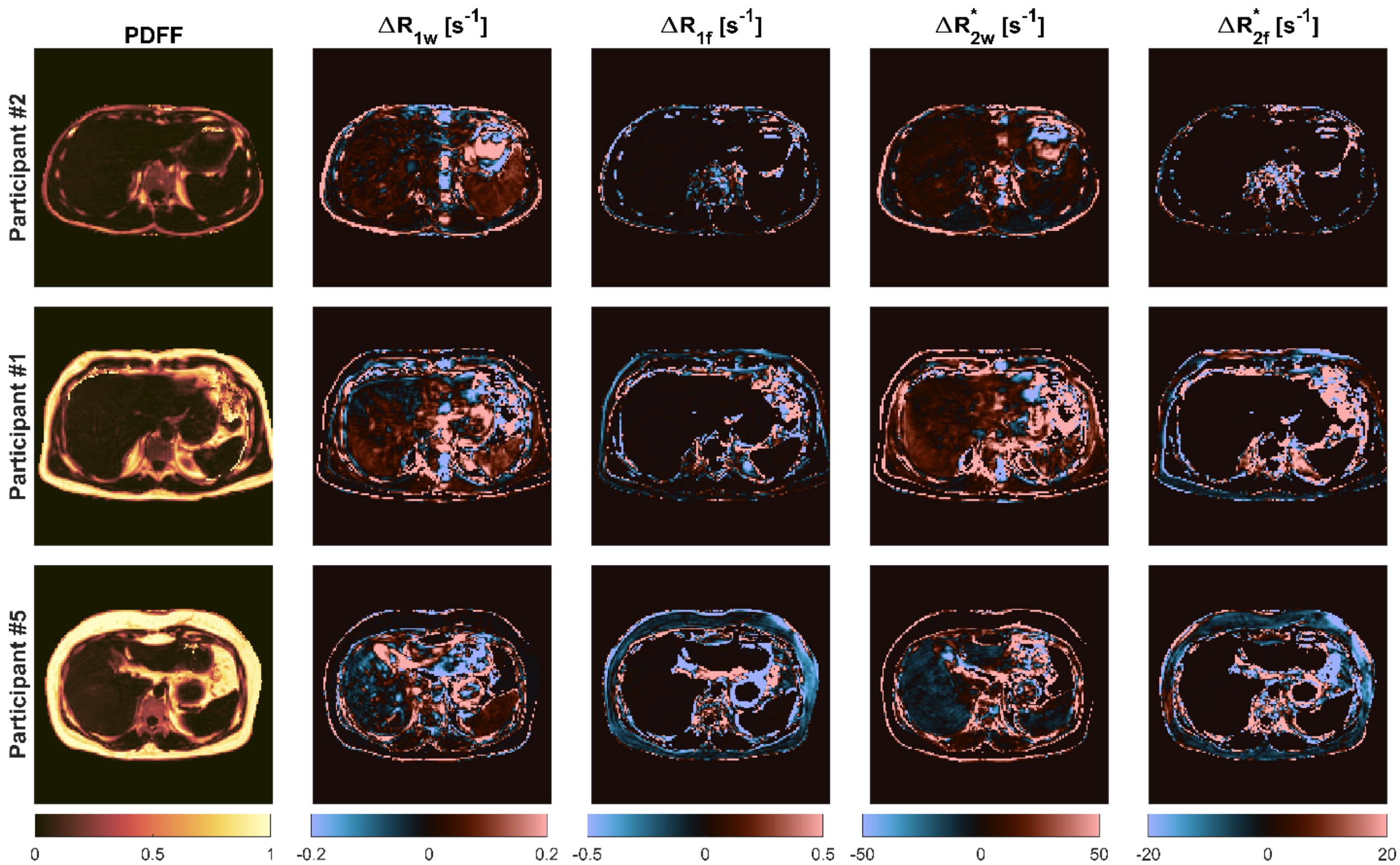

**Figure 7:** Comparison of $\Delta R_{1w}$, $\Delta R_{1f}$, $\Delta R^*_{2w}$, $\Delta R^*_{2f}$ across participants with different amounts of subcutaneous (SAT) and visceral adipose tissue (VAT). Participant #2 (top row) had relatively low SAT and VAT, precluding reliable quantification of fat-specific metrics. Participant #1 (middle row) presented more adipose tissue than the cohort average. However, motion and ringing artifacts limited accurate quantification of fat-specific metrics, particularly in the anterior abdomen and visceral fat regions. In this case, quantification was only feasible within a ROI placed in posterior SAT. Participant #5 (bottom row) exhibited sufficient SAT to enable robust visualization and quantification of fat-specific metrics.

## Discussion

MR-oximetry relies on the sensitivity of relaxation times to changes in dissolved oxygen in tissue and blood[5,6,10,16,17,42,43]. This work proposes a framework named SR-Fat DESPOT for the joint quantification of baseline MRI parameters as well as changes in the relaxation parameters induced by a gas challenge. We also developed an approach for optimizing sequence parameter selection based on CRB theory. An open-source abdominal numerical phantom was modified to evaluate the proposed methodology. Most importantly, SR-Fat DESPOT was tested in vivo to demonstrate feasibility and sensitivity.

By adopting an SR-VFA scheme[11–13] and modelling the relaxation changes induced by oxygenation within the signal model, we demonstrate that for each time frame, SR-Fat DESPOT quantifies oxygenation-induced changes with two baseline measurements and a additional single MGRE acquisition during the time frame. The proposed approach preserves quantitative capability while reducing data acquisition requirements, compared to conventional approaches that perform full data acquisition before and during the gas challenge[14–17]. Furthermore, SR-Fat DESPOT relaxes acquisition constraints by enabling the joint estimation of longitudinal and transverse relaxation rates, unlike studies reported in literature that require separate acquisitions for $\mathrm{R_1}$ and $\mathrm{R_2^*}$[5,14,17,42].

Importantly, SR-Fat DESPOT allows for the quantification of fat- and water-specific parameters while accounting for $\mathrm{R_2^*}$ effects. This capability is particularly valuable for the assessment of pathologies in organs such as the liver[44], muscle[45–47], and spleen[48,49], where $\mathrm{R_{1w}}$ has demonstrated diagnostic relevance. Prior studies have also highlighted the potential of fat-specific $\mathrm{R_1}$ metrics for assessing oxygenation changes[6,7].

The trends in measured $\Delta\mathrm{R_{1w}}$ observed in Experiments 1 and 2 for water-dominant tissues are consistent with another study at 3 T[10]. This study reported the largest response in the spleen, followed by the renal cortex, with muscle exhibiting the smallest change. We observed responses in the same order of magnitude across organs. However, quantitative differences were noted. Our measured maximum $\Delta\mathrm{R_{1w}}$ in the spleen (0.086 $s^{-1}$) exceeded a previously reported value (0.068 $s^{-1}$), whereas our measurement for the kidneys was slightly lower (0.052 $s^{-1}$ vs. 0.056 $s^{-1}$). These discrepancies could in part be from methodological differences, including the use of single-component $\Delta R_1$ in the literature versus $\Delta R_{1w}$ in our study, and our combined consideration of both renal parenchymal layers (cortex and medulla). Importantly, Experiment 2 demonstrated statistically significant increases in $\Delta R_{1w}$ for the spleen and kidneys relative to zero and between two time frames (without and with 100% $O_2$). This indicates that, despite differences in absolute values, the changes measured with SR-Fat DESPOT in these organs are driven by the hyperoxic gas challenge.

In the participant from Experiment 1, liver $\Delta R_{1w}$ and $\Delta R^*_{2w}$ were significantly different from zero regardless of the hyperoxic gas challenge. Work in literature has reported that 100% $O_2$ inhalation can increase $\Delta R_1$ while decreasing $\Delta R^*_{2w}$ in the liver, although $\Delta R^*_{2w}$-based measurements were highly variable across measurements and individuals[42]. In contrast, $\Delta R^*_{2w}$ was observed to increase in Experiment 1. This discrepancy may reflect confounding physiological effects, as $\Delta R^*_{2w}$ is sensitive to BOLD-related hemodynamic changes and may have been influenced by the breath-hold response rather than tissue oxygenation alone. This is particularly relevant because the breath-hold duration used in this study was approximately twice that used in the literature[42]. The variability observed in Experiment 1 is also consistent with the reported variability of liver $\Delta R^*_{2w}$ measurements and contrasts with Experiment 2, where no significant liver $\Delta R^*_{2w}$ changes were detected in the participants.

Our work supports the idea that $\Delta R_1$ is a suitable metric for organs with low fat content, such as the spleen and liver in healthy individuals. However, our numerical simulations demonstrate that this metric becomes biased in the presence of adipose tissue, as observed in the simulated vertebral body and kidneys in regions containing both fat and water signal components, resulting in erroneous quantification of tissue oxygenation changes.

The precision of $\Delta R_{1w}$, $\Delta R_{1f}$, $\Delta R^*_{2w}$, and $\Delta R^*_{2f}$ is critical for detecting oxygenation-induced changes. To improve sensitivity, we leveraged CRB theory and investigated the use of bipolar readout gradients. CRB theory was used to optimize sequence parameters and minimize the expected variance of the estimates[24]. Bipolar readouts enabled the acquisition of more echoes with shorter $\Delta TE$, which improves fat-water separation performance[50–55], without increasing scan time. In simulations, this improved both accuracy and precision, while in vivo it reduced artifacts in the quantitative maps. Additionally, the use of bipolar gradients reduced the IQR of $\Delta R_{1w}$ and $\Delta R^*_{2w}$ estimates.

Limitations of this study should be acknowledged. First, motion management relied on ~30 seconds breath-hold acquisitions. Although breath-hold strategies have been used in similar studies[10,42], prolonged breath-holds may limit clinical applicability. Scan times could be further reduced using parallel imaging. However, this study prioritized signal-to-noise ratio to maximize sensitivity to oxygenation changes. Second, the study evaluated a limited subset of organs under 100% $O_2$ inhalation. The physiological response depends on the gas mixture used[16,17]. Inhalation of 100% oxygen induces vasoconstriction, whereas carbogen may elicit stronger oxygenation changes[16,17] and could be explored in future work. Third, study and interpretability of $\mathrm{R_{1f}}$ and $\Delta\mathrm{R}^*_{2f}$ in vivo was limited due to low amount of adipose tissue in participants. This limitation could be addressed in future work considering a larger cohort.

**Conclusion**

SR-Fat DESPOT combines a multi-echo SR-VFA acquisition scheme with a signal model that explicitly captures oxygenation-induced changes in relaxation rates. The framework demonstrated accurate parameter estimation in numerical simulations and detected hyperoxia-induced changes in vivo, particularly in the spleen and kidneys, consistent with values reported in the literature.

**Data availability statement**

Source code for SR-Fat DESPOT, including the modifications used for the numerical abdominal phantom, is available upon request. The code requires a working installation of the ISMRM Fat-Water Separation Toolbox. In vivo data from one participant is also available upon request.

**Acknowledgments**

The authors acknowledge the developers of the ISMRM fat-water toolbox (http://www.ismrm.org/workshops/FatWater12/data.htm), and members of the MR Methods Research Group (McGill University) for useful discussion. Jorge Campos Pazmiño acknowledges funding provided from the *Fonds de recherche du Québec – Nature et technologies* (FRQNT; B2X - Bourse de doctorat en recherche). This project was funded by a Discovery Grant from the Natural Science and Engineering Research Council, *Chercheur boursier* and *Établissement de jeune* *chercheur* awards from the Fond de recherche du Québec – Santé, and by the Montreal General Hospital Foundation.

**Supplementary Information**

Supplementary information for: *Advancing oxygen-enhanced MRI acquisition and modelling through dynamic ΔR1 and ΔR2* sensing*, Jorge Campos Pazmiño, Amiel Beausoleil-Morrison, Nathalie Panov, Véronique Fortier, and Ives R. Levesque (2026)

**Supplementary Theory: Derivation of Fisher Information Matrix and Cramér-Rao Lower bounds**

Labelling the total number of flip angles (FAs) collected before and during the gas challenge as $P_{air}$ and $P_G$, respectively; and the total number of FAs $P = P_{air} + P_G$ the probability density function of the measurements is:

$$\Pr(\mathbf{S}|\mathbf{\Theta}) = \prod_{p=1}^{P}\prod_{n=1}^{N}\frac{1}{\sqrt{2\pi\sigma_S^2}}e^{-\frac{1}{2\sigma_S^2}(S_{n,p}-\bar{S})^2} = \frac{1}{\sqrt{(2\pi)^{NP}\sigma_S^{2NP}}}e^{-\frac{1}{2\sigma_S^2}\|\mathbf{S}-\bar{\mathbf{S}}\|^2} \quad \text{S1}$$

where N is the number of echoes, $\Pr(\mathbf{S}|\mathbf{\Theta})$ is the probability of the measurement **S** (vector containing all the data points) given $\mathbf{\Theta}$ (vector containing all the unknown parameters in the signal equation), $\bar{\mathbf{S}}$ is the expectation value of the data points, and $\sigma_S^2$ is the noise variance.

The entries of the Fisher Information Matrix (FIM), defined as the expectation value of the second derivative of the log-likelihood function[1], are (after setting the derivatives of constants equal to zero):

$$[\mathrm{FIM}]_{ij} = -\langle\frac{\partial}{\partial\Theta_i}\frac{\partial}{\partial\Theta_j}\ln(\Pr(\mathbf{S}|\mathbf{\Theta}))\rangle = -\langle\frac{\partial}{\partial\Theta_i}\frac{\partial}{\partial\Theta_j}\left(-\frac{1}{2\sigma_S^2}\|S-\bar{\mathbf{S}}\|^2\right)\rangle \quad \text{S2}$$

The term $\bar{\mathbf{S}}$ can be expressed as $\bar{\mathbf{S}} = A\rho$, where A and ρ correspond to:

$$A = \begin{bmatrix} e^{-R_{2w}^*TE_1}W_1\cos(\phi_{w,1} + 2\pi\psi_1 TE_1) & e^{-R_{2f}^*TE_1}W_1\sum_q \alpha_q\cos(\phi_{f,1} + 2\pi(\psi_1 + \Delta f_q)TE_1) \\ e^{-R_{2w}^*TE_1}W_1\sin(\phi_{w,1} + 2\pi\psi_1 TE_1) & e^{-R_{2f}^*TE_1}W_1\sum_q \alpha_q\sin(\phi_{f,1} + 2\pi(\psi_1 + \Delta f_q)TE_1) \\ \vdots & \vdots \end{bmatrix} \quad \text{S3}$$

$$\rho = M_0\begin{bmatrix} 1-\mathrm{PDFF} \\ \mathrm{PDFF} \end{bmatrix} \quad \text{S4}$$

The matrix A depends on the number of data points collected before and during the gas challenge. W and F correspond to the magnitude of Equations 2 and 3 in the main manuscript. For acquisitions performed before the gas challenge, $\Delta R_{1w}$, $\Delta R_{1f}$, $\Delta R_{2w}^*$, $\Delta R_{2f}^*$ are set to zero, while for acquisition during the gas challenge these variables are set as free variables. The number of rows in A depend on the number of echoes and FAs collected, with the total number of rows being equal to $NP_{air} + NP_G$.

By choosing $\boldsymbol{\Theta} = \{M_0, PDFF, R_{1w}, R_{1f}, R^*_{2w}, R^*_{2f}, \phi_{w,p}, \phi_{f,p}, \psi_p, \phi, \epsilon, \Delta R_{1w}, \Delta R_{1f}, \Delta R^*_{2w}, \Delta R^*_{2f}\}$ (a total of 12+3P unknowns), the entries of the FIM can be derived with a similar mathematical formalism previously reported in literature[2]. Thus, the entries of the FIM are:

For $i = 1,2$ and $j = 1,2$

$$[FIM]_{ij} = \frac{1}{\sigma_s^2}\frac{\partial \rho^t}{\partial \Theta_i} A^t A \frac{\partial \rho}{\partial \Theta_j} \tag{S5}$$

For $i = 1,2$ and $j = 3, \dots, 12 + 3P$

$$[FIM]_{ij} = \frac{1}{\sigma_s^2}\frac{\partial \rho^t}{\partial \Theta_i} A^t \frac{\partial A}{\partial \Theta_j} \rho \tag{S6}$$

For $i = 3, \dots, 12 + 3P$ and $j = 3, \dots, 12 + 3P$

$$[FIM]_{ij} = \rho^t \frac{1}{\sigma_s^2}\frac{\partial A^t}{\partial \Theta_i}\frac{\partial A}{\partial \Theta_j} \rho \tag{S7}$$

The equivalent number of signals averaged (NSA) is calculated as:

$$NSA_i = NP\frac{CRB(\Theta_i|\Theta_{j\neq i})}{[FIM^{-1}]_{ii}} \tag{S8}$$

where $CRB(\Theta_i|\Theta_{i\neq j})$ is the variance of $\Theta_k$ when it is the only free parameter in the model. $CRB(\Theta_i|\Theta_{i\neq j})$ can be caculated as the reciprocal of the diagonal entries of the FIM or equivalently with the following formulas:

$$CRB(\Theta_1|\Theta_{j\neq 1}) = \frac{\sigma_s^2}{\sum_{p=1}^{P}\sum_{n=1}^{N}\left\|(1-PDFF)W_p e^{i\phi_{w,p}} e^{-R^*_{2w}TE_n} + (PDFF)F_p e^{i\phi_{f,p}} \sum_{q=1}^{Q}\alpha_q e^{i2\pi\Delta f_q TE_n}\right\|^2 e^{(-1)^n 2\varepsilon}} \tag{S9}$$

$$CRB(\Theta_2|\Theta_{j\neq 2}) = \frac{\sigma_s^2}{\sum_{p=1}^{P}\sum_{n=1}^{N}\left\|-M_0 W_p e^{i\phi_{w,p}} e^{-R^*_{2w}TE_n} + M_0 F_p e^{i\phi_{f,p}} \sum_{q=1}^{Q}\alpha_q e^{i2\pi\Delta f_q TE_n}\right\|^2 e^{(-1)^n 2\varepsilon}} \tag{S10}$$

$$CRB(\Theta_3|\Theta_{j\neq 3}) = \frac{\sigma_s^2}{\sum_{p=1}^{P}\sum_{n=1}^{N}\left(M_0(1-PDFF)\frac{dW_p}{dR_{1w}} e^{-R^*_{2w}TE_n} e^{(-1)^n\varepsilon}\right)^2} \tag{S11}$$

$$CRB(\Theta_4|\Theta_{j\neq 4}) = \frac{\sigma_s^2}{\sum_{p=1}^{P}\sum_{n=1}^{N}\left(M_0(PDFF)\frac{dF_p}{dR_{1f}} e^{-R^*_{2f}TE_n} e^{(-1)^n\varepsilon}\right)^2 \left\|\sum_{q=1}^{Q}\alpha_q e^{i2\pi\Delta f_q TE_n}\right\|^2} \tag{S12}$$

$$CRB(\Theta_5|\Theta_{j\neq 5}) = \frac{\sigma_s^2}{\sum_{p=1}^{P}\sum_{n=1}^{N}\left(TE_n M_0(1-PDFF)W_p e^{-R^*_{2w}TE_n} e^{(-1)^n\varepsilon}\right)^2} \tag{S13}$$

$$CRB(\Theta_6|\Theta_{j\neq 6}) = \frac{\sigma_s^2}{\sum_{p=1}^{P}\sum_{n=1}^{N}\left(TE_n M_0(PDFF)F_p e^{-R^*_{2f}TE_n} e^{(-1)^n\varepsilon}\right)^2 \left\|\sum_{q=1}^{Q}\alpha_q e^{i2\pi\Delta f_q TE_n}\right\|^2} \tag{S14}$$

$$CRB(\Theta_{7+l}|\Theta_{j\neq 7+l}) = \frac{\sigma_s^2}{\sum_{n=1}^{N}\left(M_0(1-PDFF)W_p e^{-R^*_{2w}TE_n} e^{(-1)^n\varepsilon}\right)^2} \tag{S15}$$

$$CRB(\Theta_{7+P+p}|\Theta_{j\neq 7+P+p}) = \frac{\sigma_s^2}{\sum_{n=1}^{N}\left(M_0(PDFF)F_p e^{-R^*_{2f}TE_n} e^{(-1)^n\varepsilon}\right)^2 \left\|\sum_{q=1}^{Q}\alpha_q e^{i2\pi\Delta f_q TE_n}\right\|^2} \tag{S16}$$

$$CRB(\Theta_{7+2P+p}|\Theta_{j\neq 7+2P+p}) = \frac{\sigma_s^2}{\sum_{n=1}^{N}(2\pi TE_n)^2\|S_{n,p}\|^2} \tag{S17}$$

$$\mathrm{CRB}\left(\Theta_{7+3P+p}|\Theta_{j\neq 7+3P+p}\right)=\frac{\sigma_s^2}{\sum_{n=1}^{N}\left\|S_{n,p}\right\|^2} \qquad \text{S18}$$

$$\mathrm{CRB}\left(\Theta_{7+4P+p}|\Theta_{j\neq 7+4P+p}\right)=\frac{\sigma_s^2}{\sum_{n=1}^{N}\left\|S_{n,p}\right\|^2} \qquad \text{S19}$$

$$\mathrm{CRB}\left(\Theta_{7+5P+1}|\Theta_{j\neq 7+5P+1}\right)=\frac{\sigma_s^2}{\sum_{p=P_{air}+1}^{P}\sum_{n=1}^{N}\left(M_0(1-\mathrm{PDFF})\frac{dW_p}{dR_{1w}}e^{-R_{2w}^*TE_n}e^{(-1)^n\varepsilon}\right)^2} \qquad \text{S20}$$

$$\mathrm{CRB}\left(\Theta_{7+5P+2}|\Theta_{j\neq 7+5P+2}\right)=\frac{\sigma_s^2}{\sum_{p=P_{air}+1}^{P}\sum_{n=1}^{N}\left(M_0(\mathrm{PDFF})\frac{dF_p}{dR_{1f}}e^{-R_{2f}^*TE_n}e^{(-1)^n\varepsilon}\right)^2\left\|\sum_{q=1}^{Q}\alpha_q e^{i2\pi\Delta f_q TE_n}\right\|^2} \qquad \text{S21}$$

$$\mathrm{CRB}\left(\Theta_{7+5P+3}|\Theta_{j\neq 7+5P+3}\right)=\frac{\sigma_s^2}{\sum_{p=P_{air}+1}^{P}\sum_{n=1}^{N}\left(TE_nM_0(1-\mathrm{PDFF})W_pe^{-R_{2w}^*TE_n}e^{(-1)^n\varepsilon}\right)^2} \qquad \text{S22}$$

$$\mathrm{CRB}\left(\Theta_{7+5P+4}|\Theta_{j\neq 7+5P+4}\right)=\frac{\sigma_s^2}{\sum_{p=P_{air}+1}^{P}\sum_{n=1}^{N}\left(TE_nM_0(\mathrm{PDFF})F_pe^{-R_{2f}^*TE_n}e^{(-1)^n\varepsilon}\right)^2\left\|\sum_{q=1}^{Q}\alpha_q e^{i2\pi\Delta f_q TE_n}\right\|^2} \qquad \text{S23}$$

**Supplementary Methods: Parameters for sequence optimization and numerical simulations**

Table S1 presents the parameters considered for sequence parameter optimization and Monte Carlo (MC)-based validation of NSA calculations. Unless otherwise stated, all parameters were selected to be representative of tissue parameters and MRI acquisitions at 3 T. All calculations were performed assuming infinite readout bandwidth (rBW), rendering chemical shift- and field inhomogeneity–induced spatial misregistration negligible[3]. The minimum TR was calculated as $2TE_1 + (N-1)\Delta TE$ with the multiplicative factor 2 used to simulate a more realistic time between the last echo and the start of a new TR. Finally, for all calculations, it is assumed that $B_1^+ = 1$ with zero uncertainty.

**- Sequence parameter optimization**: $R_{1w}$ represents spleen[4] (0.76 $s^{-1}$), liver[5] (1.01 $s^{-1}$), and paraspinal muscle[4] (1.13 $s^{-1}$) and $R_{1f}$ represents subcutaneous fat[6]. $R_{2w}^*$ was selected to represent spleen[4] (30 $s^{-1}$), liver[7], and muscle[4] (50 $s^{-1}$ is a reasonable value for both structures). $R_{2f}^*$ for spleen was selected equal to $R_{2w}^*$ while for liver a different value was selected based on literature[8]. For muscle, the same difference as in liver was keep to analyze a scenario where $R_{2w}^* \neq R_{2f}^*$, even though this difference might not be realistic in muscle. $\phi_{w,p}$ and $\phi_{f,p}$ were set to zero under the assumption of perfect spoiling[9]. $\psi_p$, $\phi_p$, and $\epsilon_p$ were assumed to be the same across FAs and selected based on literature[2,10]. $\Delta R_{1w}$ and $\Delta R_{1f}$ were selected to represent the changes induced in spleen and subcutaneous fat due to 100% $O_2$ inhalation[4,11]. The spleen was selected because its response to 100% $O_2$ inhalation is consistently reported as statistically significant in the literature[4,11]. $\Delta R_{1f}$ was retrieved from MRI at 1.5 T literature due to lack of reporting for 3 T. $\Delta R_{2w}^*$ and $\Delta R_{2f}^*$ were set to 0 since no change is expected in a hyperoxic gas challenge[4,11]. The FAs $\theta_{b1}$ and $\theta_{b2}$ correspond to the datasets collected during air inhalation. The FA $\theta_t$ corresponds to a single acquisition during 100% $O_2$ inhalation. $TE_1$ and $\Delta TE$ were selected based on the minimum achievable values in our experiments such that breath holds for abdominal imaging had a reasonable duration. The number of echoes correspond to the maximum number of echoes that could be fit in our experiments with bipolar readout gradient pulses.

- **MC simulations to validate NSA calculations:** We used MC simulations with 5000 noisy realizations to estimate the variance of each parameter and to compute MC-derived NSA values using the normalization described in Supplementary Information equations S9-S23. These MC-derived NSA values were compared with those obtained from CRB theory. The same parameters and assumptions considered for sequence parameter optimization were used. Simulations were used for parameters representative of the spleen. Results for the comparison is presented in Figure S2.

**Table S1:** Summary of parameters for sequence parameter optimization and MC simulations. Values considered in each case are reported as single values or ranges in brackets (discretized using a fixed step size reported in parenthesis, i.e. [min, max] (step size)).

| | **Sequence parameter optimization** | **MC simulations to validate NSA calculations** |
|---|---|---|
| **$M_0$ [a.u.]** | 100 | 100 |
| **PDFF** | [0.00,1.00] (0.25) | [0.00,1.00] (0.1) |
| **$R_{1w}$ [$s^{-1}$]** | 0.76, 1.01, 1.13 | 0.76 |
| **$R_{1f}$ [$s^{-1}$]** | 2.62 | 2.62 |
| **$R^*_{2w}$ [$s^{-1}$]** | 30, 50 | 30 |
| **$R^*_{2f}$ [$s^{-1}$]** | 30, 56 | 30 |
| **$\phi_{f,p}$ [rad]** | 0 | 0 |
| **$\phi_{f,p}$ [rad]** | 0 | 0 |
| **$\psi_p$ [Hz]** | 40 | 40 |
| **$\phi_p$ [rad]** | 0.02π | 0.02 π |
| **$\epsilon_p$ [a.u.]** | 0.03 | 0.03 |
| **$\Delta R_{1w}$ [$s^{-1}$]** | 0.05 | 0.05 |
| **$\Delta R_{1f}$ [$s^{-1}$]** | 0.07 | 0.07 |
| **$\Delta R^*_{2w}$ [$s^{-1}$]** | 0 | 0 |
| **$\Delta R^*_{2f}$ [$s^{-1}$]** | 0 | 0 |
| **$\theta_{b1}$ [°]** | [3,30] (2) | 5 |
| **$\theta_{b2}$ [°]** | [3,30] (2) | 30 |
| **$\theta_t$ [°]** | [3,30] (2) | 30 |
| **$TE_1$ [ms]** | 1.2 | 1.2 |
| **$\Delta TE$ [ms]** | 0.9 | 0.9 |
| **TR [ms]** | $2TE_1 + (N-1)\Delta TE$ | $2TE_1 + (N-1)\Delta TE$ |
| **# echoes** | 10 (bipolar) | 10 (bipolar) |

**Supplementary Methods: Numerical phantom**

Table S2 lists the $T_1$ values incorporated into the abdominal numerical phantom[12] for all modeled organs. Values are provided to mimic both baseline conditions (room air) and a hyperoxic gas challenge (100% $O_2$), along with their corresponding standard deviations. These standard deviations were used to introduce "texture" into the phantom, consistent with the approach used in the original model for magnetic susceptibility and $R_2^*$. Specifically, texture was generated by assigning each voxel within an organ a value of $T_1 + a_t\sigma_{T_1}$, where $T_1$ and $\sigma_{T_1}$ are given in Table S2. We set $a_t$=0.3 to approximately match the variation of the magnetic susceptibility with $R_2^*$. Note that $a_t$ could have been set to 0 since the focus of the experiments with the numerical model was to evaluate the effect of noise in the quantification of the parameters. Ground truth values for single-component $R_1$ and $\Delta R_1$ were derived from noiseless data generated at FAs $\theta_{b1}$ and $\theta_{b2}$ using the theory presented in the section "Fitting for a single-component signal equation" of this document.

**Table S2:** Summary of $T_1$ and standard deviations assigned to organs modelled in the abdominal numerical phantom[12]. For each organ, the table includes a reference for the studies considered to retrieve each value.

| Organ | Baseline $T_1$ [ms] | Std. dev. $\sigma_{T_1}$ [ms] | 100% $O_2$ $T_1$ [ms] | Std. dev. $\sigma_{T_1}$ [ms] |
|---|---|---|---|---|
| **Aorta**[13] | 1649 | 68 | 1649 | 68 |
| **Inferior vena cava (IVC)**[13] | 1649 | 68 | 1649 | 68 |
| **Right kidney (RK)**[4] | 1240 | 130 | 1171 | 121 |
| **Left kidney (LK)**[4] | 1240 | 130 | 1171 | 121 |
| **Liver**[6] | 941 | 136 | 922 | 127 |
| **Spleen**[4] | 1316 | 74 | 1221 | 62 |
| **Gallbladder (GB)** | 0 | 0 | 0 | 0 |
| **Esophagus** | 0 | 0 | 0 | 0 |
| **Stomach** | 0 | 0 | 0 | 0 |
| **Pancreas**[6] | 725 | 71 | 725 | 71 |
| **Small intestine (SI)** | 0 | 0 | 0 | 0 |
| **Large intestine (LI)** | 0 | 0 | 0 | 0 |
| **Hepatic veins (HV)**[13] | 1649 | 68 | 1649 | 68 |
| **Porta vein (PV)**[13] | 1649 | 68 | 1649 | 68 |
| **Ribs**[14] | 246 | 18 | 246 | 18 |
| **Vertebrae**[6] | 586 | 73 | 586 | 73 |
| **Intervertebral disks (IVD)**[15] | 1168 | 18 | 1168 | 18 |
| **Spinal cord (SC)**[15] | 993 | 47 | 993 | 47 |
| **Subcutaneous fat (SF)**[6] | 382 | 13 | 372 | 13 |
| **Visceral fat (VF)**[6] | 382 | 13 | 372 | 13 |
| **Heart**[16] | 1125 | 55 | 1125 | 55 |
| **Muscle**[4] | 884 | 52 | 885 | 47 |
| **Air** | 0 | 0 | 0 | 0 |

**Supplementary Methods: Schematic representation of Experiments 1 and 2**

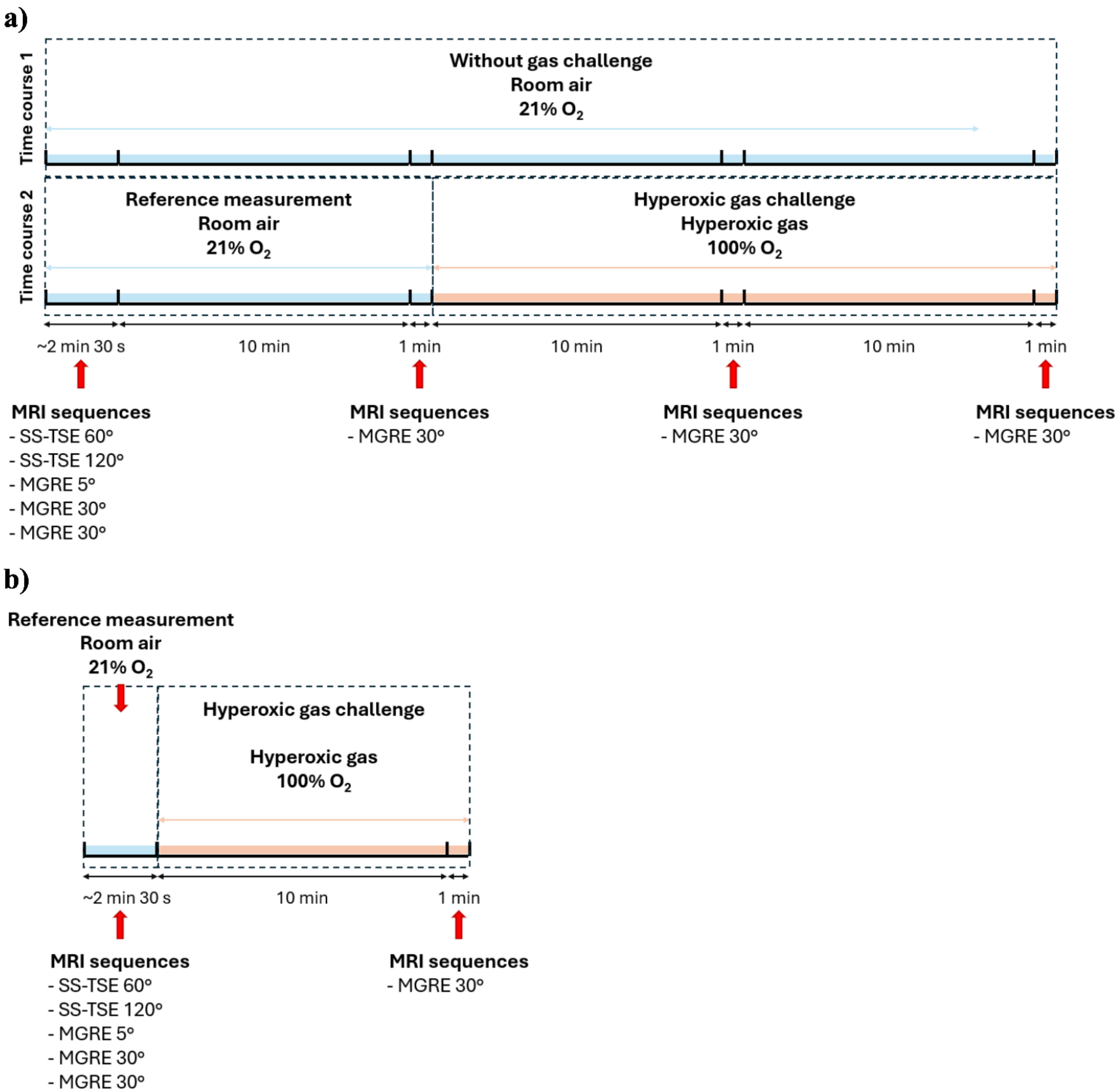

**Figure S1:** Schematic representation of in vivo experiments. a) Time course for Experiment 1 (test-retest experiment). The experiment has two different time courses. First measurement (control) is performed without a gas challenge and subsequently the experiment is repeated with a gas challenge. The two time courses were repeated twice in the same healthy participant on different days. b) Time course for Experiment 2 (multi-participant experiment). Each participant was subjected to a 10 minutes hyperoxic gas challenge.

**Supplementary Methods: Organ segmentation in vivo**

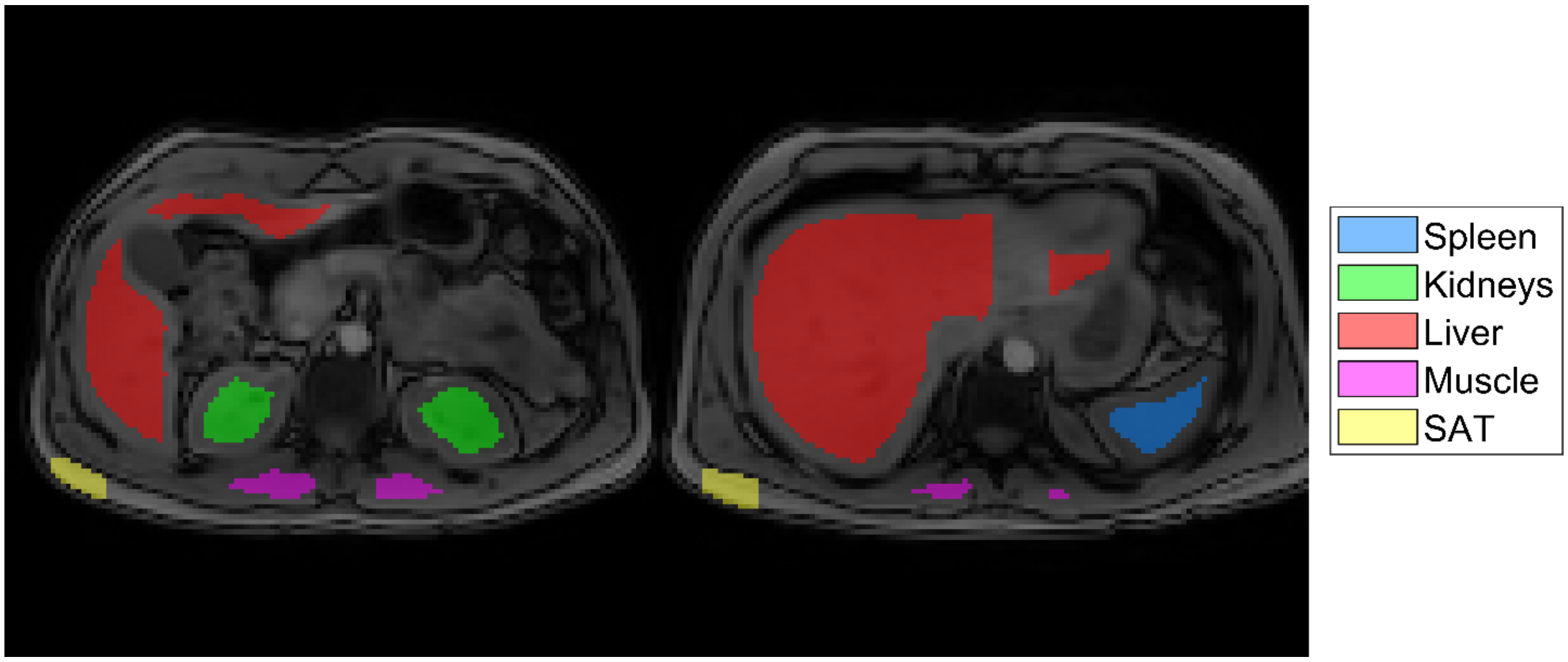

**Figure S2:** Example of automatic segmentation from Participant #1 for spleen, kidneys, liver, and muscle. Portions of masks that cover regions with pulsatile ghosting artifacts are eliminated from the segmentation results and subsequently masks are eroded (3 voxels). SAT is segmented semiautomatically by selecting voxels with PDFF>0.6 in manually selected portion of SAT.

**Supplementary Methods: Derivation of equations for initial guess for $\Delta R_{1w}$ and $\Delta R_{1f}$**

In the proposed model, during a gas challenge, at a time t, $\Delta R_{1w}$ and $\Delta R_{1f}$ are estimated from two baseline measurements and a single MGRE measurement acquired with a FA labeled as $\theta_t$. Assuming that the baseline $M_0$, $R_{1w}$, and $R_{1f}$ remained unchanged during the gas challenge and they are known from the baseline measurements, the magnitude of the water- and fat-specific signals corresponding to the data collected with excitation FA $\theta_t$, and excluding $R^*_{2w}$, $R^*_{2f}$, and $e^{(-1)^n \epsilon_p}$ effects, are $M_0(1-\mathrm{PDFF})W_t$ and $M_0(\mathrm{PDFF})F_t$ for:

$$W_t = \frac{\left(1 - e^{-(R_{1w}+\Delta R_{1w})TR}\right)\sin(B_1^+\theta_t)}{1 - e^{-(R_{1w}+\Delta R_{1w})TR}\cos(B_1^+\theta_t)} \quad \text{S24}$$

$$F_t = \frac{\left(1 - e^{-(R_{1f}+\Delta R_{1f})TR}\right)\sin(B_1^+\theta_t)}{1 - e^{-(R_{1f}+\Delta R_{1f})TR}\cos(B_1^+\theta_t)} \quad \text{S25}$$

From these equations, it is possible to solve for $R_{1w} + \Delta R_{1w}$ and $R_{1f} + \Delta R_{1f}$:

$$\Delta R_{1w} + R_{1w} = -\frac{1}{TR}\ln\left(\frac{W_t - M_0(1-\mathrm{PDFF})\sin(B_1^+\theta_t)}{W_t\cos(B_1^+\theta_t) - M_0(1-\mathrm{PDFF})\sin(B_1^+\theta_t)}\right) \quad \text{S26}$$

$$\Delta R_{1f} + R_{1f} = -\frac{1}{TR}\ln\left(\frac{F_t - M_0(\mathrm{PDFF})\sin(B_1^+\theta_t)}{F_t\cos(B_1^+\theta_t) - M_0(\mathrm{PDFF})\sin(B_1^+\theta_t)}\right) \quad \text{S27}$$

Subtracting $R_{1w}$ and $R_{1f}$ to both sides of Equations S26 and S27, respectively, Equations 5 and 6 of the main manuscript are obtained.

**Supplementary Methods: Fitting for a single-component signal equation**

For numerical simulations and in vivo experiments, we compared single- and two-component signal models for quantifying baseline parameters and gas-challenge-induced changes. This section describes the data-fitting approach employed for a single-component signal equation.

During a gas challenge, the steady-state MGRE signal magnitude in a voxel, assuming a single-component tissue model and the use of bipolar readout gradient pulses, can be modelled as[4]:

$$S_{n,p} = M_0 \frac{\left(1 - e^{-(R_1+\Delta R_1)TR}\right)\sin\left(B_1^+\theta_p\right)}{1 - e^{-(R_1+\Delta R_1)TR}\cos\left(B_1^+\theta_p\right)} e^{-(R_2^*+\Delta R_2^*)TE_n} e^{(-1)^n\epsilon_p} \qquad \text{S28}$$

where the constant $M_0$ is the total magnetization at thermal equilibrium, TR is the repetition time, $B_1^+$ represents the effect of transmit field heterogeneities, $\theta_p$ is the excitation FA, and $TE_n$ is the echo time. The total longitudinal relaxation rate is expressed as $R_1 + \Delta R_1$, where $R_1$ denotes the baseline relaxation rate measured in the absence of any gas challenge (for which $\Delta R_1 = 0$). The term $\Delta R_1$ represents the change induced by the gas challenge. An analogous notation is used for the total effective transverse relaxation rate $R_2^* + \Delta R_2^*$, where $R_2^*$ is the baseline value and $\Delta R_2^*$ is the change induced by the gas challenge. The term $\epsilon_p$ represents the amplitude modulation induced by bipolar readout gradient pulses[3,10,17,18]. In the case of a unipolar readout, this term becomes zero.

In this work, baseline $M_0$, $R_1$ and $R_2^*$ were estimated using two MGRE measurements acquired at two different FAs. Hereafter, the FAs used for baseline measurements will be labelled as $\theta_{b1}$ and $\theta_{b2}$ ("b" for "baseline"). Baseline $R_1$ was estimated by applying DESPOT1[19] independently to each echo and then averaging the resulting $R_1(TE_n)$ values across echoes. We assumed that $\epsilon_p$ remains the same across different FAs. For bipolar readouts, $R_2^*$ was obtained via mono-exponential fitting applied separately to odd and even echoes and then results were averaged between the two sets. For unipolar readouts, mono-exponential fitting was performed with all echoes.

During the gas challenge, at a time t, $\Delta R_1$ and $\Delta R_2^*$ were estimated from the two baseline measurements and a single MGRE measurement acquired with a FA labeled as $\theta_t$. For a single component case, the same derivation performed for Equations S26 or S27 can be followed and a formula for $\Delta R_1(TE_n)$ at each echo time can be derived. This derivation results in:

$$\Delta R_1(TE_n) = -\frac{1}{TR}\ln\left(\frac{S_{n,t} - M_0\, e^{-(R_2^*+\Delta R_2^*)TE_n} e^{(-1)^n\epsilon_t}\sin(B_1^+\theta_t)}{S_{n,t}\cos(B_1^+\theta_t) - M_0 e^{-(R_2^*+\Delta R_2^*)TE_n} e^{(-1)^n\epsilon_t}\sin(B_1^+\theta_t)}\right) - R_1(TE_n) \qquad \text{S29}$$

The average across echo times of Equation S29 is considered as the $\Delta R_1$ estimated for the single component case and it is calculated as:

$$\Delta R_1 = \frac{1}{N}\sum_{n=1}^{N}\left[-\frac{1}{TR}\ln\left(\frac{S_{n,t} - M_0\, e^{-(R_2^*+\Delta R_2^*)TE_n} e^{(-1)^n\epsilon_t}\sin(B_1^+\theta_t)}{S_{n,t}\cos(B_1^+\theta_t) - M_0 e^{-(R_2^*+\Delta R_2^*)TE_n} e^{(-1)^n\epsilon_t}\sin(B_1^+\theta_t)}\right) - R_1(TE_n)\right] \qquad \text{S30}$$

The total transverse relaxation $R_2^* + \Delta R_2^*$ was estimated using the same approach as for the baseline parameter, but with the data acquired at FA = $\theta_t$.

## Supplementary Results: Validation of NSA calculations using MC simulations

Using MC simulations, we validated NSA derived from CRB theory. Figure S3 demonstrates that the noise efficiency of the estimates is accurately described by CRB theory for fat- and water-specific metrics at high and low nominal proton density fat fraction (PDFF) values, respectively. Figure S3 presents a comparison of NSA values derived from CRB theory and MC simulations for multiple parameters. As show by the figure, for $M_0$ and PDFF both methods to calculate NSA agree except at nominal PDFF=0 or 1. For water- and fat-specific parameters excellent agreement exists for PDFF $\leq 0.6$ and PDFF $\geq 0.2$. Differences between both calculation methods at high and low nominal PDFF are explained due to SNR reduction for water- and fat-specific parameters, which introduces bias in their estimation (*i.e.* breaking the assumptions of CRB theory).

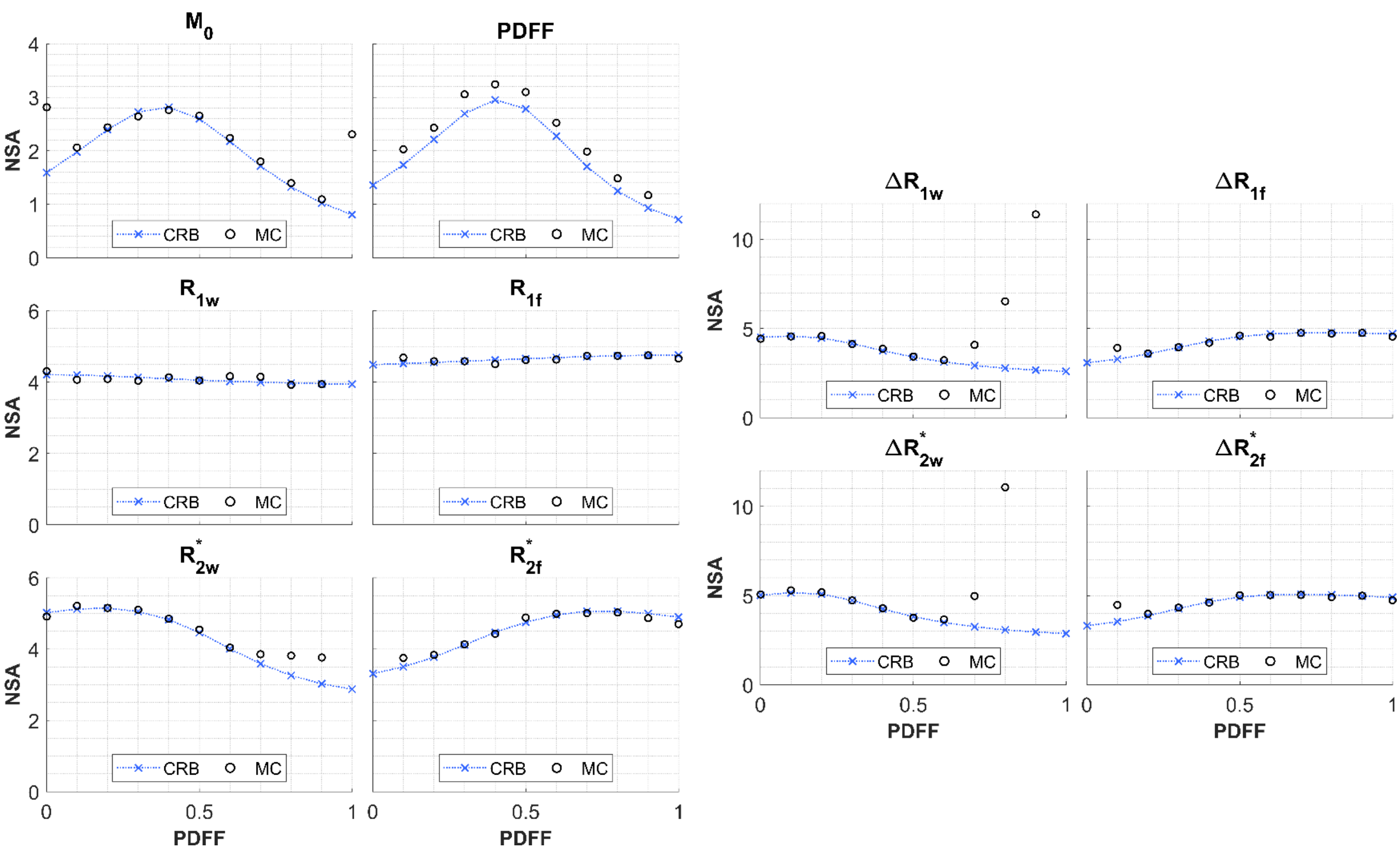

**Figure S3**: Comparison of NSA values derived from CRB theory (blue line) and MC numerical simulation (black markers). Comparison is presented for $M_0$, PDFF, $R_{1w}$, $R_{1f}$, $R_{2w}^*$, $R_{2f}^*$, $\Delta R_{1w}$, $\Delta R_{1f}$, $\Delta R_{2w}^*$, and $\Delta R_{2f}^*$. The terms $\phi_{w,p}$, $\phi_{f,p}$, $\psi_p$, $\phi_p$, $\epsilon_p$ were also included in both, CRB and MC-based derivations, but are omitted from the figure.

## Supplementary Results: Comparison of constrained and unconstrained FA optimization for SR-Fat DESPOT

We performed FA optimization using CRB theory with the constraint $\theta_t = \theta_{b2}$ and without it (i.e. $\theta_t \neq \theta_{b2}$). For the constrained optimization, the optimal FA pair $(\theta_{b1}, \theta_{b2} = \theta_t)$ was $(5°, 30°)$, which prioritizes noise performance for $\Delta R_{1w}$. In contrast, when baseline $R_{1w}$ and $R_{1f}$, were prioritized, the optimal pair was $(30°, 5°)$.

For optimization allowing $\theta_t \neq \theta_{b2}$, FA combinations ($\theta_{b1}$, $\theta_{b2}$, $\theta_t$) that maximized noise performance for $\Delta R_{1w}$ and $\Delta R_{1f}$ were (5°, 14°, 30°) and (7°, 23°, 3°), respectively. However, these FAS reduce performance for baseline parameter estimation compared with FA selections constrained by$\theta_{b2} = \theta_t$.

Figure S4 compares NSA across PDFF values for FAs $(5°, 30°)$, $(30°, 5°)$, (5°, 14°, 30°), and (7°, 23°, 3°). Averaged across PDFF, the pair $(5°, 30°)$ provides 25% higher NSA for $\Delta R_{1w}$ than $(30°, 5°)$ , confirming its advantage for $\Delta R_{1w}$ noise performance. Although results for $\theta_t \neq \theta_{b2}$ (5°, 14°, 30°) and (7°, 23°, 3°) optimize NSA for $\Delta R_{1w}$and $\Delta R_{1f}$, respectively, this improvement came at the expense of baseline parameter performance. Specifically, averaged across PDFF, $(5°, 30°)$ provides 14% and 54% higher NSA for $R_{1w}$and $R_{1f}$, respectively, than (5°, 14°, 30°); and 38% higher NSA for $R_{1f}$ than (7°, 23°, 3°). Overall, the FA pair $(5°, 30°)$ provided the most balanced performance.

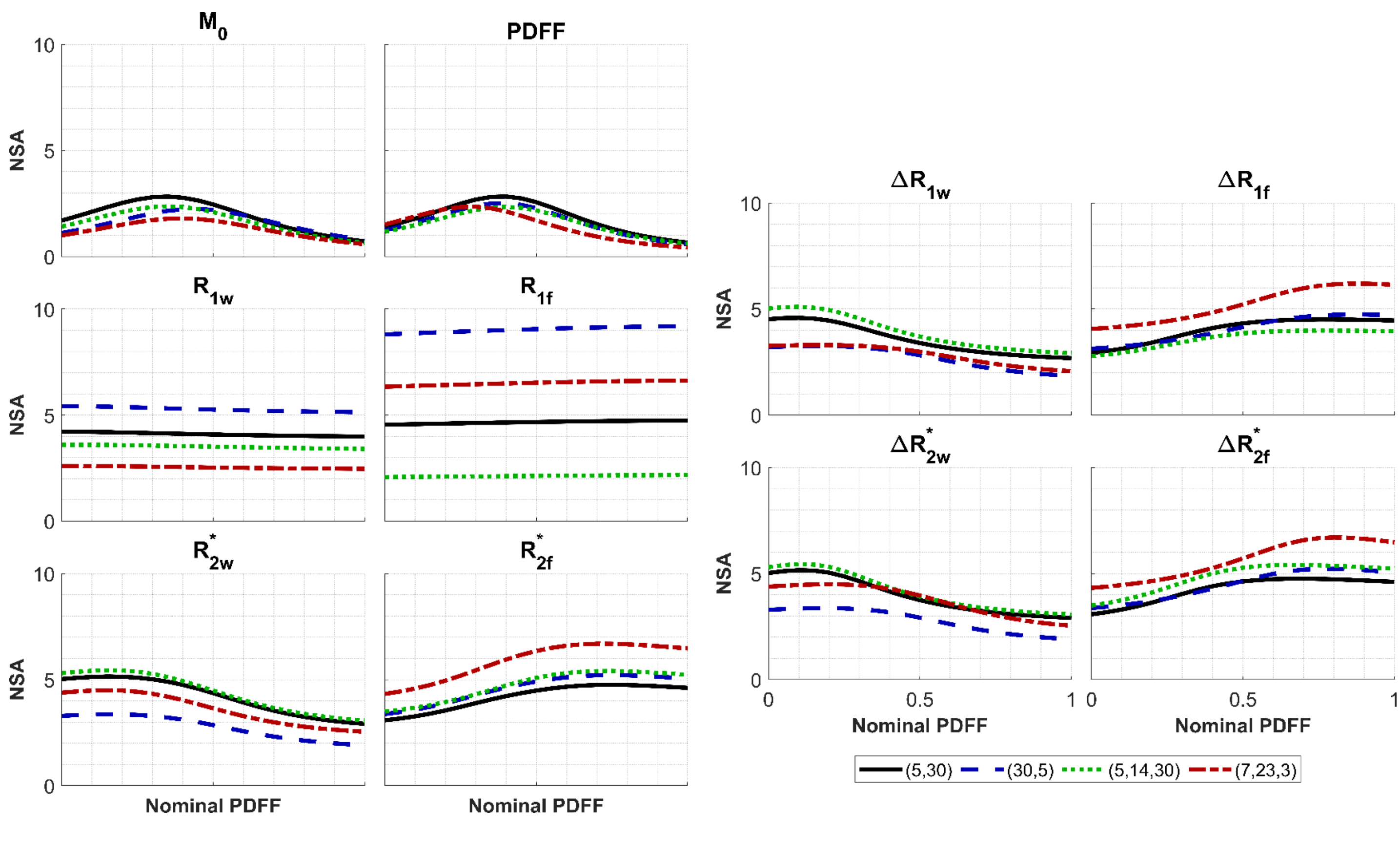

**Figure S4:** NSA curves as function of FA for varying nominal PDFF ($\phi_{f,p}$, $\psi_p$, $\phi_p$, $\epsilon_p$ were included in calculations but are omitted from the figure). These curves compare optimization with the constrained $\theta_t = \theta_{b2}$ and without it (i.e. $\theta_t \neq \theta_{b2}$). The pairs (5°, 30°) and (30°, 5°) are the optimal options for constrained optimization and prioritize performance for $\Delta R_{1w}$ and baseline parameters ($R_{1w}$and $R_{1f}$), respectively. The sets (5°, 14°, 30°), and (7°, 23°, 3°) are the optimal. options to maximize performance for $\Delta R_{1w}$ and $\Delta R_{1f}$, respectively. These sets degrade the performance for $R_{1w}$and $R_{1f}$ compared to (5°, 30°) and (30°, 5°).

**Supplementary Results: Numerical phantom with simulated unipolar gradients**

Figures S5 and S6 present results obtained from the numerical model with both phase error and amplitude modulation set to zero. These figures replicate the calculations shown in Figures 3 and 4 of the main manuscript, but with the numerical phantom modeled using unipolar readout gradient pulses.

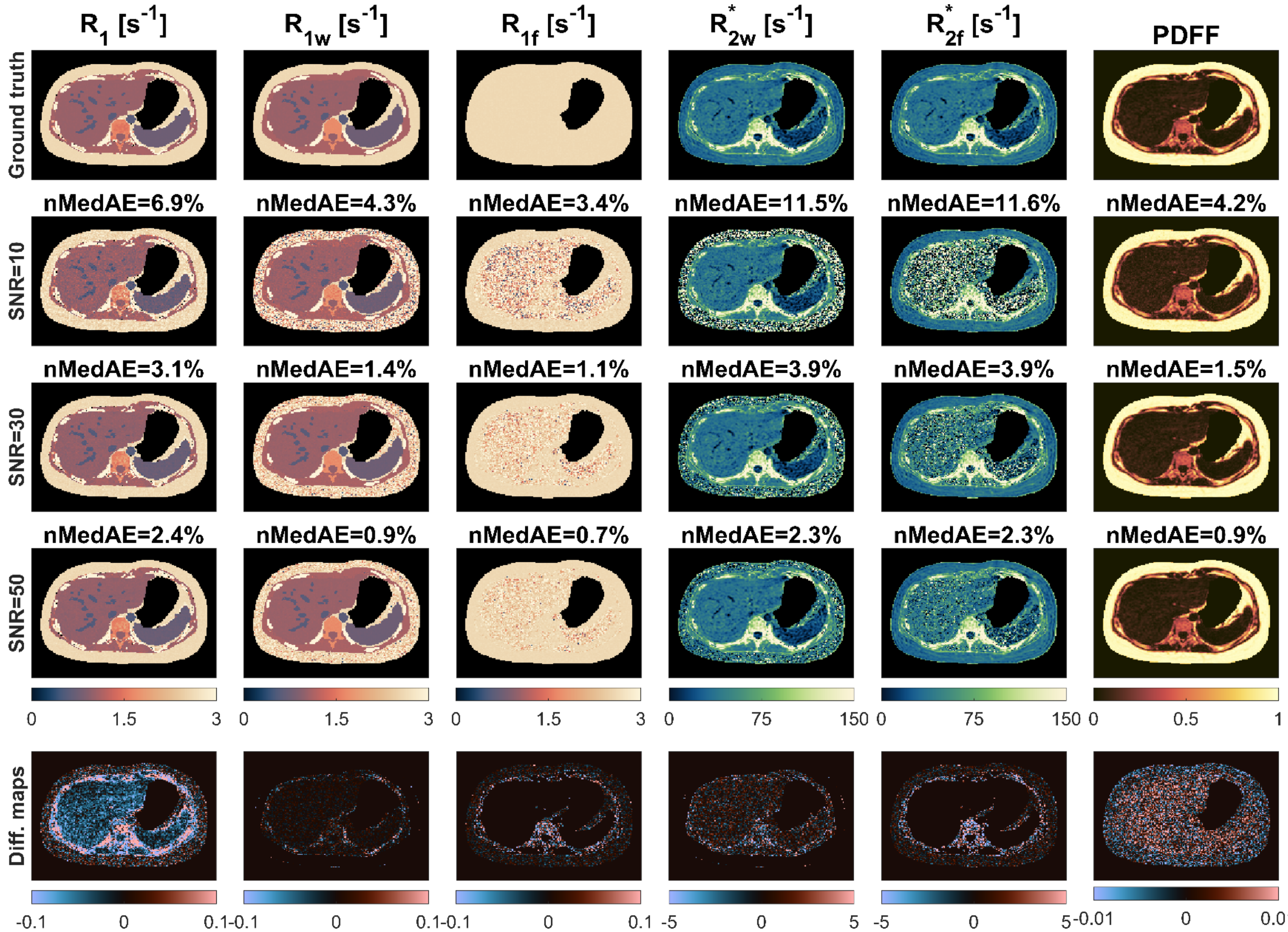

**Figure S5:** Comparison of baseline quantitative maps and ground truth for $R_1$, $R_{1w}$, $R_{1f}$, $R^*_{2w}$, $R^*_{2f}$, and PDFF using the numerical phantom modelled with unipolar readout gradients. Rows (top to bottom) show the ground truth from the numerical simulation followed by reconstructed results at different SNR levels (10, 30, and 50). The final row presents difference maps relative to the ground truth for SNR = 50. Columns correspond to the different model parameters. The text above each map reports the normalized median absolute error (nMedAE) computed between the ground truth and the estimated maps.

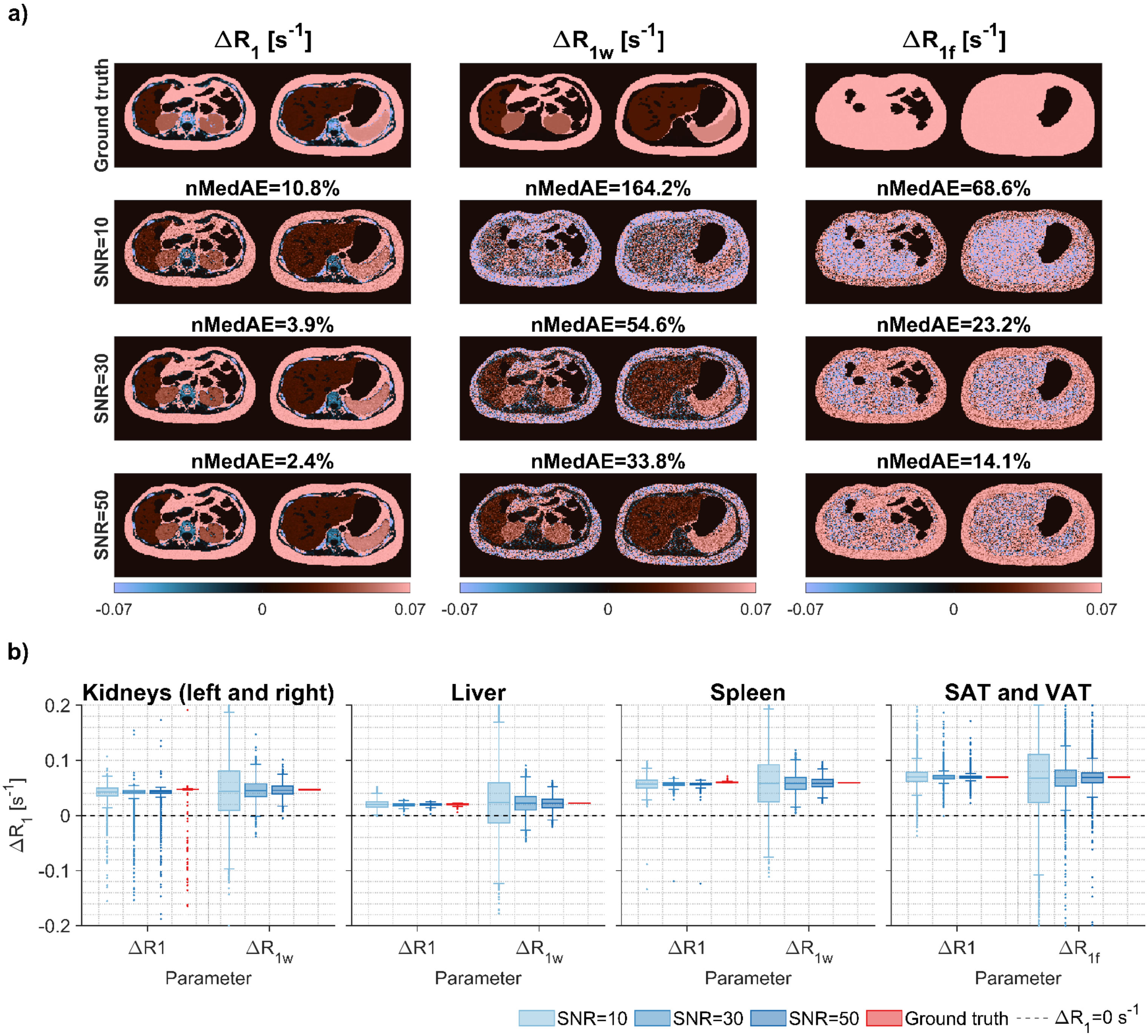

**Figure S6:** a) Comparison of quantitative maps for changes induced by a hyperoxic gas challenge (100% $O_2$ inhalation) and ground truth for $\Delta R_1$, $\Delta R_{1w}$ and $\Delta R_{1f}$ using the numerical phantom modelled with unipolar readout gradients. The text above each map reports the nMedAE computed between the ground truth and the estimated maps, b) Box plots summarizing $\Delta R_1$, $\Delta R_{1w}$ and $\Delta R_{1f}$ evaluated in masks covering the kidneys (left and right), liver, spleen, and subcutaneous and visceral adipose tissue. Box plots display the median and quartiles, while the whiskers extend to 1.5 times the interquartile range. Results are presented at different SNR levels (SNR=10, 30, and 50).

**Supplementary Results: $\Delta R^*_{2w}$ and $\Delta R^*_{2f}$ for the numerical phantom**

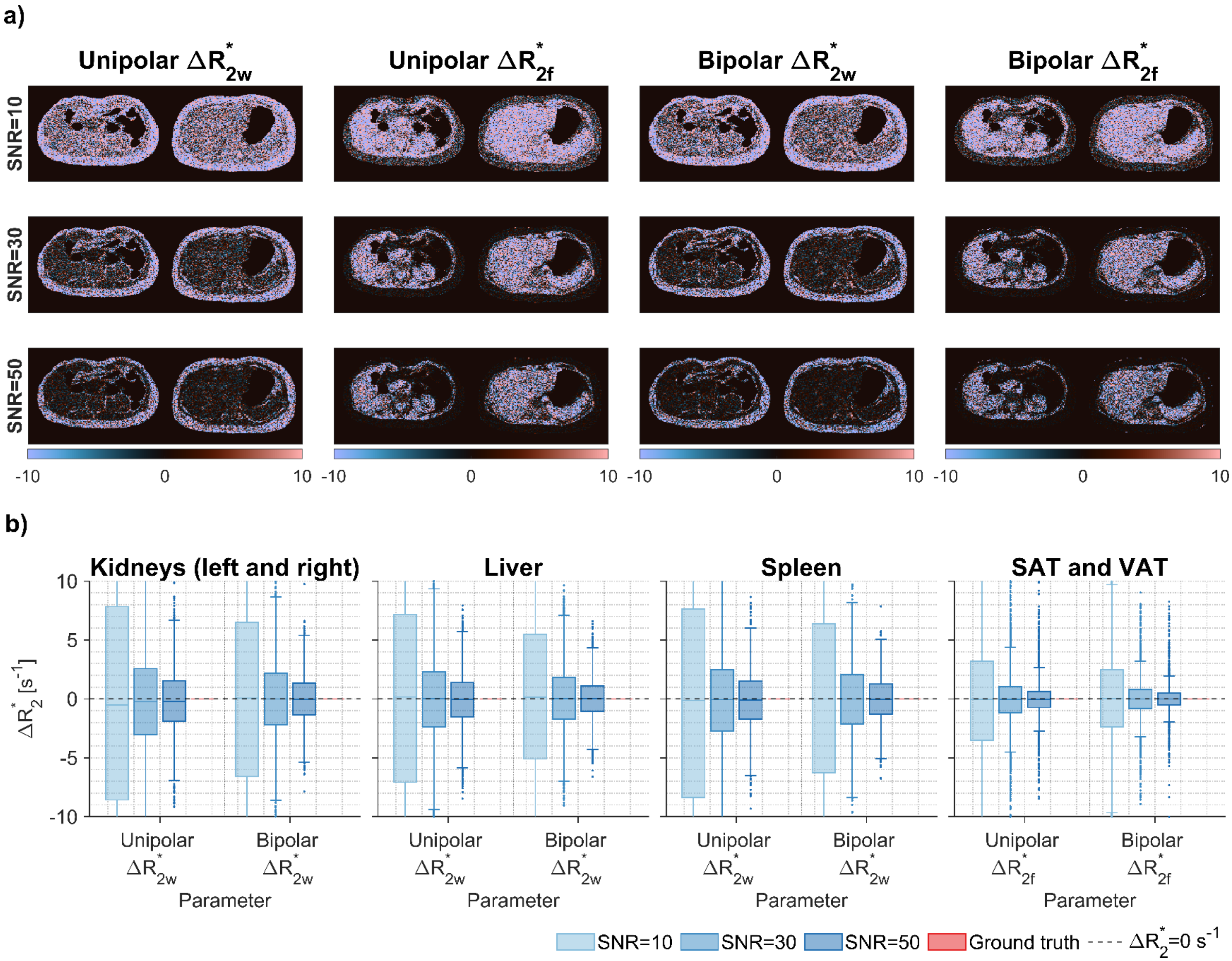

**Figure S7:** a) Comparison of quantitative maps for changes induced by a hyperoxic gas challenge (100% $O_2$ inhalation) for $\Delta R^*_{2w}$ and $\Delta R^*_{2f}$ using the numerical phantom modelled with unipolar and bipolar readout gradients. b) Box plots summarizing $\Delta R^*_{2w}$ and $\Delta R^*_{2f}$ for unipolar and bipolar gradients evaluated in masks covering the kidneys (left and right), liver, spleen, and subcutaneous and visceral adipose tissue. Box plots display the median and quartiles, while the whiskers extend to 1.5 times the interquartile range. Results are presented at different SNR levels (SNR=10, 30, and 50).

**Supplementary Results: Comparison of $\Delta R_{1w}$ and $\Delta R^*_{2w}$ in the two repetitions of Experiment 1**

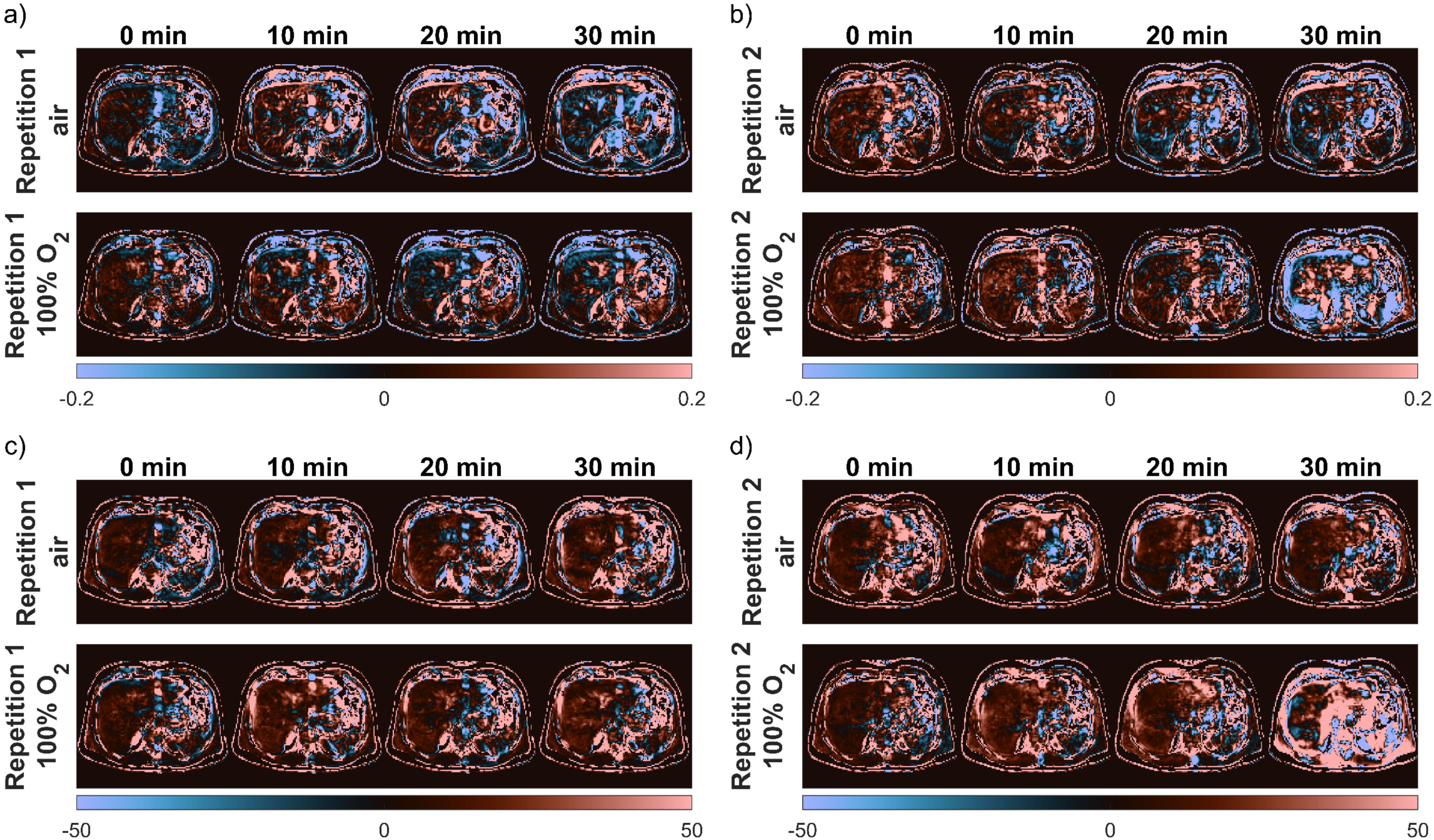

**Figure S8:** Results for Experiment 1 (test-retest experiment). Quantitative $\Delta R_{1w}$ and $\Delta R^*_{2w}$ maps at 0, 10, 20, and 30 minutes are shown for two time courses acquired without and with 100% $O_2$ inhalation. a) and b) present $\Delta R_{1w}$ results from two separate repetitions. c) and d) present $\Delta R^*_{2w}$ results from two separate repetitions. In Repetition 2, data acquired during the gas challenge at 30 minutes were corrupted due to failed motion management during the breath-hold.

**Supplementary References**